\documentclass[12pt]{iopart}
\expandafter\let\csname equation*\endcsname\relax
\expandafter\let\csname endequation*\endcsname\relax
\usepackage{amsmath,amsthm,amsfonts,amssymb,amscd,bm}
\usepackage{graphicx} 
\usepackage{multirow,booktabs}
\usepackage[table]{xcolor}
\usepackage{fullpage}
\usepackage{lastpage}
\usepackage{enumitem}
\usepackage{fancyhdr}
\usepackage{mathrsfs}
\usepackage{wrapfig}
\usepackage{setspace,caption}
\usepackage{calc}
\usepackage{multicol}
\usepackage{cancel}
\usepackage{cite}
\usepackage[retainorgcmds]{IEEEtrantools}
\usepackage[margin=3cm]{geometry}
\usepackage{empheq}
\usepackage{framed}
\usepackage[most]{tcolorbox}
\usepackage{xcolor}
\usepackage{fancyhdr}
\usepackage{hyperref}

\eqnobysec

\colorlet{shadecolor}{orange!15}

\renewcommand{\r}{{\mathbf r}}

\renewcommand{\r}{\mathbf r}

\renewcommand{\P}{\mathbb{P}}

\begin{document}

\title[Approach to Steady-state in Nested Stochastic Resetting: Accumulation times]{Approach to Steady-State in Nested Resetting Processes}

\author{Callum Britton$^1$, Ben Le Jeune$^1$, Henry Alston$^{1,2}$,  Paul C. Bressloff$^1$, Thibault Bertrand$^{1,*}$}
\address{$^1$\it{Department of Mathematics, Imperial College London, 180 Queen’s Gate, London SW7 2AZ}}
\address{$^2$\it{Laboratoire de physique de l'\'Ecole normale sup\'erieure,
		CNRS, PSL University, Sorbonne Universit\'e, and Universit\'e 
		Paris-Cit\'e, 75005 Paris, France}}
\ead{t.bertrand@imperial.ac.uk}
\vspace{10pt}
\begin{indented}
\item[]X 20XX
\end{indented}

\begin{abstract}
We characterise the approach to nonequilibrium steady state in nested resetting processes by deriving accumulation times for the system, which quantify the effective first-passage time for the local establishment of steady state. For equal resetting rates, we obtain closed-form expressions showing that relaxation propagates as a wavefront with a delay that increases linearly along the resetting chain. We then extend this analysis to heterogeneous resetting rates, deriving exact steady-state distributions, spatial moments and accumulation times for both degenerate and non-degenerate resetting rates. We show that relaxation to steady state is ultimately governed by the minimum resetting rate in the system and, in degenerate systems, its multiplicity. These results provide a comprehensive analytical description of first-passage to nonequilibrium steady state in hierarchically coupled stochastic resetting systems, with potential applications to relaxation and transport processes in biological systems.
\end{abstract}

\section{Introduction}

Stochastic resetting is a paradigm for non-equilibrium dynamics in which a system undergoes random, instantaneous returns to a prescribed reference state, interrupting the underlying dynamics at random renewal times~\cite{Evans11a,Evans20}. Originally introduced in the context of a Brownian particle that resets to its starting position at a constant Poisson rate~\cite{Evans11a}, the framework has since been extended to encompass a wide variety of underlying processes, reset protocols, and reset distributions~\cite{Evans20,Pal16,Nagar23,Gupta22}. A hallmark of stochastic resetting is the emergence of a nonequilibrium steady state (NESS): the competition between diffusive spreading and resetting events arrests the natural tendency of the distribution to broaden, giving rise to a stationary distribution that is generically non-Gaussian and breaks detailed balance~\cite{Evans11b,Eule16,Mendez16,Bello23,Roberts24}. Alongside providing a well-defined steady-state distribution, resetting has a striking effect on dynamical properties such as first-passage times and search efficiencies~\cite{Janson12,Bray13,Evans13,Reuveni16,Pal17,Bressloff20,Besga20,Majumdar21,Faisant21,DeBruyne22,Alston25a}; in particular, stochastic resetting has recently attracted significant attention as tuning the resetting rate was shown to minimise the mean first-passage time to a target in a variety of settings~\cite{Evans11a,Reuveni16,Pal17}. While early work focused primarily on single-particle Brownian motion~\cite{Evans11a,Evans11b,Evans20}, the effect of resetting in more complex settings (including interacting particles, active matter, and network-structured systems) remains under active investigation~\cite{Nagar23,Meigel23,Bressloff24,Alston25a,Alston25b,Alston26,Biroli24,Singh20,Pal16,Georgiou26}.

Beyond the characterisation of the NESS itself, a fundamental question concerns the \emph{dynamics} of relaxation: how does a system evolve from an arbitrary initial condition towards its nonequilibrium steady state, and on what timescale? For a single Brownian particle with Poissonian resetting, this question has been studied from several complementary perspectives. Majumdar {\it et al.}~\cite{Majumdar15} showed that the approach to the NESS exhibits a sharp \emph{dynamical transition}: at any fixed time $t$, there exists a critical distance $x^*(t) \sim t$ from the reset point such that the distribution has effectively reached its stationary form for $|x| < x^*(t)$, while remaining in a transient regime for $|x| > x^*(t)$. The boundary $x^*(t)$ propagates ballistically outward, separating a central NESS region from an outer transient region. This picture is closely related to the propagation of information from the reset point outward through the system, and provides a natural geometric interpretation of relaxation. A complementary spectral approach to relaxation was developed by Bressloff~\cite{Bressloff20}, who studied the approach to the NESS via the eigenspectrum of the Fokker-Planck operator in the presence of resetting, and by Krapivsky and Redner~\cite{Krapivsky2020}, who analysed the transient distribution for diffusion with resetting in detail. More recently, the relaxation of systems with non-Poissonian resetting protocols~\cite{Pal16,Eule16} and resetting in confining potentials~\cite{Singh20,Gupta19} has been studied, revealing that the qualitative picture of a propagating NESS boundary persists across a broad class of resetting processes.

One particularly useful and physically transparent approach to characterising relaxation to steady state is that of the so-called \emph{accumulation time}, introduced in the context of morphogen gradient formation during developmental patterning~\cite{Berez10,Berez11,Gordon11}. In developmental biology, cells read out the local concentration of a diffusing morphogen molecule as a proxy for their position along the embryo axis. These concentration gradients, originating from a localised source and shaped by diffusion and degradation, are not established instantaneously: the formation of the gradient is a dynamical process that takes time, and this timescale imposes a fundamental limit on the speed of positional information transfer during patterning~\cite{Berez10,Berez11,Gordon11}. An analogous accumulation time arises in the context of intracellular protein gradients during cell division~\cite{Bressloff19}. The core idea is to treat the fractional deviation of the time-dependent concentration from its steady-state value as a cumulative distribution function for the \emph{local} accumulation time, which is the effective first-passage time for the establishment of steady state at a given spatial location. This definition naturally captures the fact that different spatial regions relax at different rates, and is closely related to the propagating NESS boundary identified by Majumdar et al.~\cite{Majumdar15}. More recently, Bressloff~\cite{Bressloff21} calculated the accumulation time for a single Brownian particle with stochastic resetting, showing that the probability density associated with trajectories that have reset at least once evolves in a manner directly analogous to a morphogen concentration gradient. In \cite{Bressloff21}, Bressloff establishes an alternative, mathematically tractable method for tracking non-equilibrium relaxation and traveling front dynamics in resetting processes without relying on saddle-point approximations or large deviation theory used in Majumdar et al.~\cite{Majumdar15}. These ideas were also applied to more general resetting protocols~\cite{Bressloff22} and to resetting processes with absorption~\cite{Bressloff20}. The accumulation time provides a single scalar function of position that encodes the full spatial structure of the relaxation dynamics.

The dynamics of resetting systems are often tractable through {\it renewal equation} approaches, which exploit the fact that the system's state at time $t$ depends only on the dynamics since the last resetting event~\cite{Evans11a,Evans20}. This insight has motivated the construction of more complex interacting-particle models with resetting-like dynamics as a route to tractable nonequilibrium steady states \cite{Alston25b,Alston26}. In the simplest many-body setting, a collection of diffusive particles subject to a \emph{global} resetting protocol (in which all particles simultaneously return to their initial positions) develops non-zero higher-order correlations despite independent motion between resets~\cite{Biroli24}. The global nature of the resetting is essential: it introduces correlations that are absent in the single-particle case.

More recently, resetting-like dynamics have been explored at the level of pairwise interactions, giving rise to a richer class of many-body nonequilibrium models. In Ref.~\cite{Alston25b}, a chain of resetting processes was introduced for $N$ Brownian particles with positions $X_i(t) \in \mathbb{R}$. It was assumed that at a fixed Poisson rate $r$, the position $X_i(t)$ (independently of the others) hops to the \emph{instantaneous} position $X_{i-1}(t)$ of the preceding particle in the chain, for $i = 2, \ldots, N$, while $X_1(t)$ resets to a fixed time-independent distribution of positions $P_0(x)$. This process, termed \emph{Nested Stochastic Resetting}, supports a nonequilibrium steady state with tractable pairwise correlations of the form $\langle X_n X_{n+j}\rangle$~\cite{Alston25b}. While steady-state properties of the model were characterised in detail in Ref.~\cite{Alston25b}, but its relaxation dynamics to the NESS are yet to be settled. A key feature, stemming from the chain-like structure, is that the approach to steady state is not expected to be homogeneous across the system: as the steady state depends on the distribution $P_0(x)$, information must propagate down the chain through the resetting-like interactions before the $i$-th particle can approach its steady
state. Characterising this propagation, and in particular quantifying the timescale on which each particle locally establishes its NESS, is the central goal of the present work.

Beyond its mathematical interest, the nested resetting model provides a tractable framework for studying information propagation in hierarchically organised stochastic systems like social networks \cite{Lazer18,Stein23}. Hierarchical signal transduction cascades are ubiquitous in signaling or developmental pathways in biological systems, from kinase phosphorylation cascades~\cite{Heinrich2002} to DNA replication or protein folding. All these systems are vulnerable to error accumulation as one moves down the hierarchical structure. In turn, life invests time and energy into mechanisms like kinetic proofreading to increase the accuracy of these sequential processes \cite{Murugan12}. Further, in such systems, the rate at which each component resets to its upstream signal may vary along the cascade, motivating the study of nested resetting with heterogeneous rates. The accumulation time then quantifies the effective delay in information propagation along the cascade. While the present work focuses on the idealised setting of Brownian particles in one dimension, the analytical results obtained here provide a foundation for understanding relaxation in more realistic hierarchical signalling models.

\begin{figure}[ht!]
\begin{center}
\includegraphics[width=0.9\linewidth]{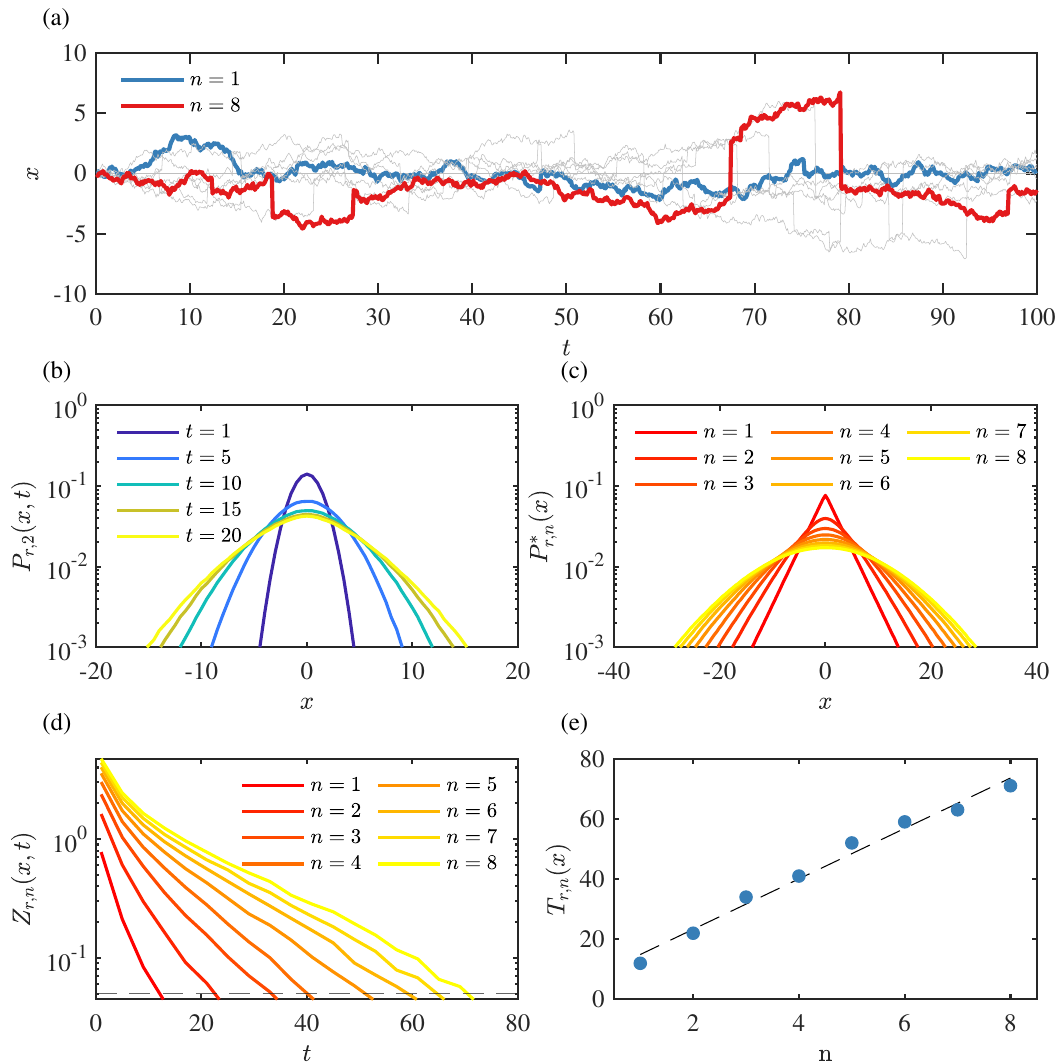}
\caption{\textit{Approach to steady state schematic.} (a) Example trajectories for a nested resetting chain of length $8$ with particles $1$ and $8$ highlighted in red and blue respectively; (b) Time-dependent distribution of particle $n=2$ plotted at a range of times, showing a widening of the distribution as the particle relaxes to steady state; (c) Steady-state distributions of particles $1$ to $8$ $(r=D=1)$; (d) Relative deviation from steady state $Z_{r,n}(x,t)$, equation~\ref{reldev}, plotted for a range of $n$ as a function of $t$, $x\approx \alpha^{-1}$; (e) Accumulation time of particle $n$, $T_{r,n}(x)$, equation~\ref{accum_Tn}, displaying a linear increase in the time taken for the system to reach steady state, $x\approx \alpha^{-1}$. All simulations performed are for $r=1,D=1$.}\label{fig:schematic} 
\end{center}
\end{figure}

The main goal of the current paper is to calculate the accumulation times characterising the propagation of steady state in nested resetting; see Figure~\ref{fig:schematic} for an illustration of the key phenomenology. We begin in Section~\ref{sec:equal} by recapping previous results on the steady-state distribution for nested resetting with equal resetting rates~\cite{Alston25b} and present a new calculation for the full time-dependent probability distribution, obtained via both a renewal-equation approach and a Fokker-Planck approach. In Section~\ref{sec:accumulation} we extend the definition of the accumulation time introduced in Ref.~\cite{Bressloff21} to the nested resetting setting. We characterise the approach to steady state first for a pair of nested resetters and then for $N$ particles with equal resetting rates, finding that relaxation propagates as a wavefront with a delay that increases \emph{linearly} along the chain, which is a direct consequence of the sequential propagation of information through the resetting hierarchy. In Section~\ref{sec:hetero}, we extend our results on the steady-state distribution, spatial moments, and accumulation times to nested resetting with heterogeneous resetting rates, analysing both degenerate and non-degenerate cases. We show that relaxation to steady state is ultimately governed by the \emph{minimum} resetting rate in the system and, in degenerate systems, by its multiplicity. We conclude in Section~\ref{sec:conclusion} with a discussion of the results and directions for future work.

\section{Nested Resetting with Equal Resetting Rates}
\label{sec:equal}

Consider nested stochastic resetting for $N$ Brownian particles on $\mathbb{R}$ with identical diffusion coefficient $D$ and initial conditions $X_i(0)=0$, $i=1,\ldots,N$. In this section we restrict our attention to the case of equal resetting rates, $r_i = r$ for all $i = 1,\ldots,N$; the general heterogeneous case is treated in Section~\ref{sec:hetero}. The position $X_{i}(t)$ resets at a Poissonian rate $r$ to the \emph{instantaneous} position of the preceding particle in the hierarchical chain:
\begin{equation}
\label{eq:langevin_i}
X_{i}(t+\mathrm{d}t)=
\begin{cases}
X_{i-1}(t+\mathrm{d}t) & \text{with probability } r\,\mathrm{d}t, \\
X_{i}(t)+\sqrt{2D\,\mathrm{d}t}\,\xi_i(t) & \text{with probability }
1-r\,\mathrm{d}t,
\end{cases}
\end{equation}
for $i=2,\ldots,N$, and
\begin{equation}
\label{eq:langevin_1}
X_{1}(t+\mathrm{d}t)=
\begin{cases}
0 & \text{with probability } r\,\mathrm{d}t, \\
X_{1}(t)+\sqrt{2D\,\mathrm{d}t}\,\xi_1(t) & \text{with probability }
1-r\,\mathrm{d}t,
\end{cases}
\end{equation}
where $\xi_i(t)$, $i=1,\ldots,N$, are independent, zero-mean, unit-variance Gaussian white noises, and $P_0(x)=\delta(x)$ is the static distribution particle $1$ resets to. The update rules~(\ref{eq:langevin_i})--(\ref{eq:langevin_1}) are understood to leading order in $\mathrm{d}t$, so that the two outcomes are mutually exclusive and exhaustive at each time step. A key feature of the update rule~(\ref{eq:langevin_i}) is that particle $i$ resets to the \emph{simultaneous} position $X_{i-1}(t+\mathrm{d}t)$ of particle $i-1$, evaluated at the \emph{same} time step. This is in contrast to simpler cascade or relay models in which particle $i$ would reset to the position $X_{i-1}(t)$ at the \emph{previous} time step. The simultaneous coupling introduces genuine many-body correlations that persist at steady state~\cite{Alston25b}, and is essential for the tractability of the model via renewal theory, since the state of particle $i-1$ at the moment of reset is itself drawn from the time-dependent distribution $P_{r,i-1}(x,t)$, that is the time-dependent distribution of position of particle $i-1$ in a nested resetting chain with resetting rate $r$.

In this section, we rederive results first presented in~\cite{Alston25b} regarding the steady-state properties of a nested resetting chain with equal resetting rates, and extend this analysis to determine the full time-dependent distribution. We conclude the section by characterising the approach to steady state via the accumulation time, {\it i.e.} the effective timescale for the local establishment of the NESS at a given point in space. Direct numerical simulations show that the accumulation time at large $x$ grows linearly across the nested resetting chain, see Figure~\ref{fig:schematic}, where the accumulation time is defined as
\begin{equation}\label{accum_Tn}
    T_{r,n}(x) = -\int_{0}^{\infty}\mathrm{d}t\; t\,\partial_t Z_{r,n}(x,t) = \int_{0}^{\infty}\mathrm{d}t\; Z_{r,n}(x,t),
\end{equation}
with
\begin{equation}\label{reldev}
    Z_{r,n}(x,t) = \frac{\bigl|P^*_{r,n}(x) - P_{r,n}(x,t)\bigr|}{P^*_{r,n}(x)}
\end{equation}
the relative deviation of the distribution of particle $n$ from its steady state distribution. We aim to analytically prove the linear increase in the accumulation time in this work.

\subsection{Renewal theory approach}
\label{subsec:renewal}

Renewal theory provides a natural and powerful framework for stochastic resetting, owing to its ability to produce a succinct integral equation for the full time-dependent probability distribution~\cite{Evans11a,Evans20}. Renewal equations come in two flavours: the \emph{last-renewal} equation, in which the PDF is constructed using information about the last time the process was renewed, and the \emph{first-renewal} equation, in which the PDF is constructed using information about the first renewal event. Throughout this section, we work with the last-renewal equation, which is better suited to deriving the steady-state distribution and the accumulation time (see Section~\ref{sec:accumulation}).

The last-renewal equation for a single particle ($n=1$) reads
\begin{equation}
\label{ren1}
P_{r,1}(x,t) = e^{-rt}\,G_{0}(x,t) + r\int_0^t \mathrm{d}\tau\, e^{-r\tau}\,G_{0}(x,\tau).
\end{equation}
The first term in equation~(\ref{ren1}) corresponds to trajectories with no reset event in $[0,t]$ — a contribution that decays to zero as $t\to\infty$ — while the second term accounts for trajectories whose last reset occurred at time $t-\tau$, after which the particle diffuses freely for a duration $\tau$ with the causal Green's function (propagator) of free diffusion,
\begin{equation}
\label{G0}
G_{0}(x,t) = \frac{1}{\sqrt{4\pi Dt}}\,\exp\!\left(-\frac{x^{2}}{4Dt}\right).
\end{equation}

Before proceeding, we record the following identity, which will be used repeatedly throughout the paper:
\begin{equation}
\label{eq:laplace_G0}
\int_0^{\infty}\mathrm{d}\tau\, e^{-r\tau}\,G_{0}(x,\tau) = \frac{1}{r}\,\cdot\frac{\alpha}{2}\,e^{-\alpha|x|} \equiv \frac{1}{r}\,P_{r,1}^{*}(x),
\end{equation}
where $\alpha = \sqrt{r/D}$. This follows from the standard Laplace transform of the Gaussian propagator~\cite{Abramowitz72} and identifies the time-integral of the weighted propagator with the single-particle steady-state distribution.

Taking the limit $t\to\infty$ in equation~(\ref{ren1}) and applying~(\ref{eq:laplace_G0}) yields the steady-state PDF
\begin{equation}
\label{ss1}
P_{r,1}^{*}(x) = \lim_{t\to\infty} P_{r,1}(x,t) = r\int_0^{\infty}\mathrm{d}\tau\, e^{-r\tau}\,G_{0}(x,\tau) = \frac{\alpha}{2}\,e^{-\alpha|x|},
\end{equation}
where $\alpha^{-1}=\sqrt{D/r}$ is the typical length diffused between consecutive resets. The steady-state distribution is a Laplace (double exponential) distribution, a hallmark of Poissonian stochastic resetting~\cite{Evans11a,Evans11b}.

We now extend the last-renewal equation to any particle in the nested chain. For equal rates, the last-renewal equation for particle $n$ reads
\begin{equation}
\label{renn}
P_{r,n}(x,t) = e^{-rt}\,G_{0}(x,t) + r\int_0^t \mathrm{d}\tau\, e^{-r\tau} \int^{+\infty}_{-\infty}\mathrm{d} \xi \;G_{0}(x-\xi,\tau)\,P_{r,n-1}(\xi,t-\tau),
\end{equation}
for $n=1,\ldots,N$, with $P_{r,0}(x,t)=P_0(x)\,\Theta(t)$, $P_0(x)=\delta(x)$, and $\Theta(t)$ the Heaviside step function. The contribution from trajectories that last reset at time $t-\tau$ now depends on the position of particle $n-1$ at time $t-\tau$, resulting in a spatial convolution between $P_{r,n-1}$ and $G_0$. The nested structure of the system, however, allows us to eliminate this dependence and rewrite equation~(\ref{renn}) in a closed form that does not explicitly involve the distribution of any other particle in the chain:
\begin{equation}
\label{renn2}
P_{r,n}(x,t) = e^{-rt}\sum_{j=0}^{n-1}\frac{(rt)^{j}}{j!}\,G_{0}(x,t) + r\int_{0}^{t}\mathrm{d}\tau\,\frac{(r\tau)^{n-1}}{(n-1)!}\, e^{-r\tau}\,G_{0}(x,\tau).
\end{equation}
This result is established by induction on $n$, using the Chapman--Kolmogorov identity for the Gaussian propagator,
\begin{equation}
\label{eq:CK}
G_0(x,t) = \int^{+\infty}_{-\infty}\mathrm{d}\xi \;G_{0}(x-\xi,\tau)\,G_0(\xi,t-\tau),
\end{equation}
The base case $n=1$ is equation~(\ref{ren1}). Assuming the result holds for $n-1$, one substitutes~(\ref{renn2}) (with $n$ replaced by $n-1$) into the convolution term of~(\ref{renn}) and applies~(\ref{eq:CK}) to collapse the resulting double convolution, yielding~(\ref{renn2}) for $n$ (see \ref{app:induction} for details of the induction proof).

An important observation is that the explicit time-dependent solution~(\ref{renn2}) admits a natural decomposition
\begin{equation}
\label{deco}
P_{r,n}(x,t) = P_{r,a_{n}}(x,t) + P_{r,d_{n}}(x,t),
\end{equation}
where
\begin{equation}
\label{Pra}
P_{r,a_{n}}(x,t) = r\int_{0}^{t}\mathrm{d}\tau\,\frac{(r\tau)^{n-1}}{(n-1)!}\, e^{-r\tau}\,G_{0}(x,\tau),
\end{equation}
is the \emph{accumulating} component and
\begin{equation}
\label{Prd}
P_{r,d_{n}}(x,t) = e^{-rt}\sum_{j=0}^{n-1}\frac{(rt)^{j}}{j!}\,G_{0}(x,t).
\end{equation}
is the \emph{decaying} component. In particular,
\begin{equation}
\label{eq:limits_a}
P_{r,a_{n}}(x,0) = 0,\qquad \lim_{t\to\infty} P_{r,a_{n}}(x,t) = P_{r,n}^{*}(x),
\end{equation}
whereas
\begin{equation}
\label{eq:limits_d}
P_{r,d_{n}}(x,0) = \delta(x),\qquad \lim_{t\to\infty} P_{r,d_{n}}(x,t) = 0.
\end{equation}
That is, $P_{r,d_{n}}(x,t)$ encodes the decay of the initial condition, while $P_{r,a_{n}}(x,t)$ encodes the approach to the unique steady state $P_{r,n}^{*}(x)$. As previously shown for single-particle resetting~\cite{Bressloff21}, the decomposition~(\ref{deco}) provides the starting point for the construction of the accumulation times quantifying relaxation to the steady state; see Section~\ref{sec:accumulation}.

\subsection{Fokker-Planck approach}
\label{subsec:FP}

The Fokker-Planck equation provides a complementary approach to the renewal framework, and is particularly convenient for deriving steady-state distributions and for handling the degenerate-rate case of Section~\ref{sec:hetero}. For particle $n$ in the nested chain, the Fokker-Planck equation reads
\begin{equation}
\label{FPn}
\partial_{t}P_{r,n}(x,t)
= D\,\nabla^2 P_{r,n}(x,t) - r\,P_{r,n}(x,t) + r\,P_{r,n-1}(x,t),
\end{equation}
with initial condition $P_{r,n}(x,0)=\delta(x)$ and $P_{r,0}(x,t)=\delta(x)\,\Theta(t)$.

\subsubsection{Steady-state distribution.} 
\begin{figure}[t!]
\begin{center}
\includegraphics[width=\linewidth]{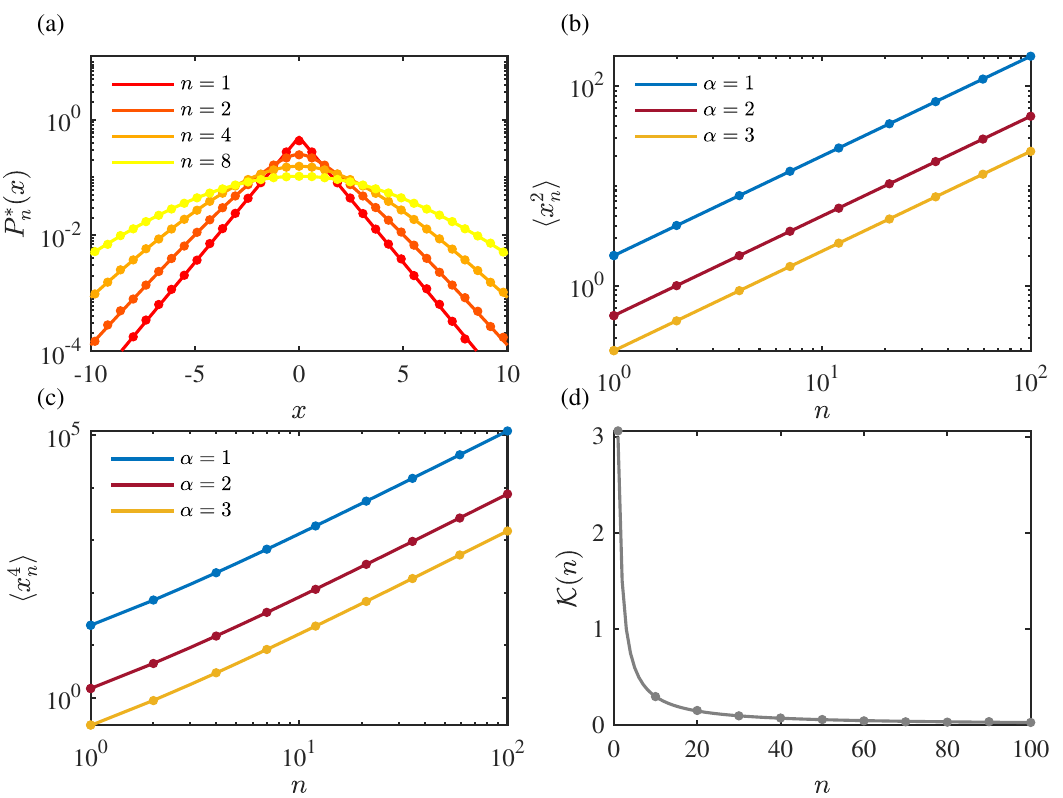}
\caption{\textit{Steady-state properties of nested resetting.} (a)~Steady-state
distributions $P^*_n(x)$ for nested resetters with equal resetting rates,
equation~(\ref{ssn}), for $n\in\{1,2,4,8\}$, resetting rate $r=1$, and
diffusivity $D=1$. (b)~Steady-state spatial variance $\langle x^2_n\rangle$
of nested resetters with equal resetting rates, equation~(\ref{xn}), for
$\alpha\in\{1,2,3\}$. (c)~Steady-state fourth spatial moment
$\langle x^4_n\rangle$ of nested resetters with equal resetting rates,
equation~(\ref{xn}), for $\alpha\in\{1,2,3\}$. (d)~Excess kurtosis
$\mathcal{K}(n)$ of nested resetters with equal resetting rates,
equation~(\ref{Kurt}). Solid lines indicate analytical results and symbols
indicate numerical simulations.}
\label{fig:ssequal}
\end{center}
\end{figure}

At steady state, $\partial_t P_{r,n}^* = 0$, and Fourier-transforming
equation~(\ref{FPn}) in space gives the recursion
\begin{equation}
\label{eq:FT_ss_recursion}
\widetilde{P}_{r,n}^{*}(k) = \frac{r}{r+Dk^{2}}\,\widetilde{P}_{r,n-1}^{*}(k), \quad n=1,\ldots,N,
\end{equation}
with $\widetilde{P}_{r,0}^{*}(k)=1$. Iterating yields the explicit solution
\begin{equation}
\label{PPn}
\widetilde{P}_{r,n}^{*}(k) = \frac{r^{n}}{\left(r+Dk^{2}\right)^{n}}, \quad n=1,\ldots,N.
\end{equation}
Since multiplication in Fourier space corresponds to convolution in real space, the steady-state PDF of particle $n$ is the $n$-fold convolution of the single-particle steady-state distribution~(\ref{ss1}) with itself. Inverting~(\ref{PPn}) using the standard Fourier transform of the modified Bessel function of the second kind~\cite{Abramowitz72}, we obtain
\begin{equation}
\label{ssn}
P_{r,n}^{*}(x) = \frac{\alpha}{\sqrt{\pi}\,(n-1)!} \left(\frac{\alpha|x|}{2}\right)^{n-1/2} K_{n-1/2}(\alpha|x|),
\end{equation}
where $K_\nu$ denotes the modified Bessel function of the second kind of order $\nu$~\cite{Abramowitz72}. As a consistency check, we verify that~(\ref{ssn}) reduces to the Laplace distribution~(\ref{ss1}) for $n=1$. Using the identity $K_{1/2}(z) = \sqrt{\pi/(2z)}\,e^{-z}$~\cite{Abramowitz72}, we find
\begin{equation}
P_{r,1}^{*}(x)
= \frac{\alpha}{\sqrt{\pi}\cdot 0!}  \left(\frac{\alpha|x|}{2}\right)^{1/2}  K_{1/2}(\alpha|x|)
= \frac{\alpha}{\sqrt{\pi}}  \left(\frac{\alpha|x|}{2}\right)^{1/2}  \sqrt{\frac{\pi}{2\alpha|x|}}\,e^{-\alpha|x|}
= \frac{\alpha}{2}\,e^{-\alpha|x|},
\end{equation}
in agreement with~(\ref{ss1}). We numerically confirm equation~(\ref{ssn}) in Figure~\ref{fig:ssequal}(a) for different $n$ values.

\subsubsection{Spatial moments.} Spatial moments are readily calculated from the characteristic function~(\ref{PPn}). Since $\widetilde{P}_{r,n}^*(k)$ is an even function of $k$, all odd moments vanish by symmetry, and the $2q$-th spatial moment reads
\begin{equation}
\label{xn}
\left\langle x_{n}^{2q}\right\rangle = (-1)^q \left.\frac{\mathrm{d}^{2q}\widetilde{P}_{r,n}^*(k)} {\mathrm{d}k^{2q}}\right|_{k=0} = \binom{n+q-1}{n-1}\frac{(2q)!}{\alpha^{2q}}.
\end{equation}
In particular, setting $q=1$ gives the variance
\begin{equation}
\label{eq:variance}
\left\langle x_n^2\right\rangle = \frac{2n}{\alpha^2} = \frac{2nD}{r},
\end{equation}
which grows linearly with $n$, consistent with the results of Refs.~\cite{Evans11b,Alston25b}. Setting $q=2$ gives the fourth moment
\begin{equation}
\label{eq:fourth_moment}
\left\langle x_n^4\right\rangle = \frac{12n(n+1)}{\alpha^4}.
\end{equation}
We numerically confirm equations~(\ref{eq:variance}) and (\ref{eq:fourth_moment}) for $q=1,2$ in Figures~\ref{fig:ssequal}(b) and~\ref{fig:ssequal}(c), respectively.

An interesting consequence of equations~(\ref{eq:variance}) and (\ref{eq:fourth_moment}) is that the excess kurtosis, a measure of non-Gaussianity, decays to zero as $n$ grows large:
\begin{equation}
\label{Kurt}
\mathcal{K}(n)= \frac{\left\langle x_{n}^{4}\right\rangle}{\left\langle x_{n}^{2}\right\rangle^{2}} - 3 = \frac{3}{n}.
\end{equation}
Note that $\mathcal{K}(1)=3$, recovering the known excess kurtosis of the Laplace distribution, and $\mathcal{K}(n)\to 0$ as $n\to\infty$, consistent with the approach to Gaussianity along the chain. This approach to Gaussianity is a consequence of the central limit theorem: the steady-state distribution of particle $n$ is the $n$-fold convolution of the single-particle Laplace distribution, and by the central limit theorem (CLT) this converges in distribution to a Gaussian as $n\to\infty$. However, we stress that the approach to Gaussianity is in the sense of the kurtosis only: the full distribution $P_{r,n}^*(x)$ is \emph{not} Gaussian for any finite $n$, as is evident from the Bessel function form~(\ref{ssn}). We numerically confirm equation~(\ref{Kurt}) in Figure~\ref{fig:ssequal}(d).

\subsubsection{Decomposed Fokker-Planck equations.} Extending the analysis of single-particle resetting~\cite{Bressloff21}, we
write down the Fokker-Planck equations for the components $P_{r,a_n}$ and
$P_{r,d_n}$ defined in~(\ref{deco})--(\ref{Prd}):
\begin{subequations}
\begin{eqnarray}
\label{Pa}
\partial_{t}P_{r,a_{n}} &=& D\,\nabla^2 P_{r,a_{n}}  - r\,P_{r,a_{n}} + r\,P_{r,a_{n-1}},  \quad P_{r,a_{n}}(x,0) = 0, \\
\label{Pd}
\partial_{t}P_{r,d_{n}} &=& D\,\nabla^2 P_{r,d_{n}}  - r\,P_{r,d_{n}} + r\,P_{r,d_{n-1}},  \quad P_{r,d_{n}}(x,0) = \delta(x),
\end{eqnarray}
\end{subequations}
where $P_{r,a_{0}}(x,t)=P_{0}(x)\,\Theta(t)$ and $P_{r,d_{0}}(x,t)=0$, preserving the nested structure. For $n>1$, $P_{r,a_{n}}$ no longer has a source at $x=0$; instead it is driven by the flux from $P_{r,a_{n-1}}$. Further, $P_{r,d_{n}}$ now has a non-zero source in $P_{r,d_{n-1}}$ for $n>1$.

As a consistency check, we integrate equations~(\ref{Pa}) and~(\ref{Pd}) over $x\in\mathbb{R}$, defining the spatially integrated quantities
\begin{equation}
\mathcal{P}_{r,a_n}(t)
\equiv \int_{-\infty}^{+\infty} P_{r,a_n}(x,t)\,\mathrm{d}x,
\qquad
\mathcal{P}_{r,d_n}(t)
\equiv \int_{-\infty}^{+\infty} P_{r,d_n}(x,t)\,\mathrm{d}x,
\end{equation}
to obtain the linear ODEs
\begin{subequations}
\begin{align}
\frac{\mathrm{d}{\cal P}_{r,a_{n}}(t)}{\mathrm{d}t}&= -r\,{\cal P}_{r,a_{n}}(t) + r\,{\cal P}_{r,a_{n-1}}(t), \\
\frac{\mathrm{d}{\cal P}_{r,d_{n}}(t)}{\mathrm{d}t}&= -r\,{\cal P}_{r,d_{n}}(t) + r\,{\cal P}_{r,d_{n-1}}(t).
\end{align}
\end{subequations}
These are linear ODEs with the boundary conditions ${\cal P}_{r,a_0}(t) = \Theta(t)$ and ${\cal P}_{r,d_0}(t) = 0$, and can be solved
iteratively to give
\begin{subequations}
\begin{align}
\label{eq:Pa_integrated}
{\cal P}_{r,a_{n}}(t) &= 1 - e^{-rt}\sum_{j=0}^{n-1}\frac{(rt)^{j}}{j!}, \\
\label{eq:Pd_integrated}
{\cal P}_{r,d_{n}}(t) &= e^{-rt}\sum_{j=0}^{n-1}\frac{(rt)^{j}}{j!}.
\end{align}
\end{subequations}
One can verify directly that ${\cal P}_{r,a_n}(t)+{\cal P}_{r,d_n}(t)=1$ for all $t\geq 0$, consistent with probability conservation, and that ${\cal P}_{r,a_n}(t)\to 1$, ${\cal P}_{r,d_n}(t)\to 0$ as $t\to\infty$, as required. Note that ${\cal P}_{r,d_n}(t)$ is the regularised incomplete gamma function $\Gamma(n,rt)/\Gamma(n)$, i.e.\ the survival function of a Gamma$(n,r)$ random variable evaluated at $t$; this connection provides an alternative probabilistic interpretation of the decay component.

\subsubsection{Alternative representation via Fourier-Laplace transform.}

The corresponding solutions of the full equations~(\ref{Pa})--(\ref{Pd}) are given by equations~(\ref{Pra}) and~(\ref{Prd}), respectively. Here, we derive an alternative representation of $P_{r,a_{n}}(x,t)$ that is particularly useful for computing the accumulation times in Section~\ref{sec:accumulation}. We proceed by performing a double Fourier-Laplace transform with respect to $x$ and $t$ defined as
\begin{equation}
\widehat{P}_{r,a_{n}}(k,s) \equiv \int^{+\infty}_{-\infty}\mathrm{d}x\, e^{-ikx} \int_{0}^{\infty}\mathrm{d}t\, e^{-st}\,P_{r,a_{n}}(x,t)
\end{equation}
to equation~(\ref{Pa}), and using the zero initial condition $P_{r,a_n}(x,0)=0$ (so there is no boundary term from the Laplace transform of the time derivative), which gives
\begin{equation}
\label{eq:FL_Pa}
s\,\widehat{P}_{r,a_{n}}(k,s) = -\left(Dk^{2}+r\right)\widehat{P}_{r,a_{n}}(k,s)  + r\,\widehat{P}_{r,a_{n-1}}(k,s).
\end{equation}
We can make use of the nested structure and iterating~(\ref{eq:FL_Pa}), we obtain
\begin{equation}
\label{bob}
\widehat{P}_{r,a_{n}}(k,s) = \frac{r}{s+Dk^{2}+r}\,\widehat{P}_{r,a_{n-1}}(k,s) = \frac{r^{n}}{s\left(s+Dk^{2}+r\right)^{n}}.
\end{equation}
where we have used the fact that
\begin{equation}
\widehat{P}_{r,a_0}(k,s)
= \frac{\widetilde{P}_{r,0}(k)}{s} = \frac{1}{s},
\end{equation}
To invert this expression, we use the partial-fraction identity
\begin{equation}
\label{eq:pf_identity}
\frac{1}{s} - \frac{(r+Dk^2)^n}{s\left(s+Dk^{2}+r\right)^{n}}= \frac{1}{s+r+Dk^2}\sum_{j=0}^{n-1}  \left(\frac{r+Dk^2}{s+r+Dk^2}\right)^{j},
\end{equation}
which is a standard geometric series identity and can be verified by multiplying both sides by $s(s+r+Dk^2)^n$. Using equation~(\ref{PPn}), identity~(\ref{eq:pf_identity}) is equivalent to
\begin{equation}
\frac{1}{s}\,\widetilde{P}_{r,n}^{*}(k)- \frac{r^{n}}{s\left(s+Dk^{2}+r\right)^{n}} = \sum_{j=0}^{n-1}  \frac{r^{j}\,\widetilde{P}_{r,n-j}^{*}(k)}{[s+r+Dk^2]^{j+1}}.
\end{equation}
Combining with equation~(\ref{bob}) and inverting the Fourier-Laplace transform, we arrive at
\begin{equation}
\label{Pra2}
P_{r,a_{n}}(x,t)= P_{r,n}^{*}(x)  - \sum_{j=0}^{n-1}\frac{(rt)^{j}}{j!}\,e^{-rt}\,    P_{r,n-j}^{*}(x)*G_{0}(x,t),
\end{equation}
where $*$ denotes a spatial convolution. The inversion of each term in the sum uses the convolution theorem together with the standard Laplace-transform pair $\mathcal{L}^{-1} \{[s+r+Dk^2]^{-(j+1)}\} = \frac{t^j}{j!}e^{-(r+Dk^2)t}$, which in real space gives $\frac{(rt)^j}{j!}e^{-rt}G_0(x,t)$ after accounting for the spatial Fourier factor.

Combining~(\ref{Pra2}) with~(\ref{Prd}), the full time-dependent PDF reads
\begin{equation}
\label{eq:full_PDF}
P_{r,n}(x,t) = e^{-rt}\sum_{j=0}^{n-1}\frac{(rt)^{j}}{j!}  \left[G_{0}(x,t) - P_{r,n-j}^{*}(x)*G_{0}(x,t)\right]  + P_{r,n}^{*}(x).
\end{equation}
This representation makes the approach to steady state transparent: as $t\to\infty$, the prefactor $e^{-rt}\sum_{j=0}^{n-1}(rt)^j/j!$ vanishes (since it equals $P_{r,d_n}(t)\to 0$, cf.\ equation~(\ref{eq:Pd_integrated})), and $P_{r,n}(x,t)\to P_{r,n}^*(x)$ as required.

\section{Approach to Steady State}
\label{sec:accumulation}
\subsection{Theory of accumulation times}

To motivate the calculation of the accumulation time of a nested resetter, we first consider a single resetting particle (corresponding to the first particle in a nested resetting chain) that resets to the origin, i.e.\ $P_0(x)=\delta(x)$. Following Ref. \cite{Bressloff21}, see also equations (\ref{Pa}) and (\ref{Pd}), we decompose the probability distribution as
$P_{r,1}(x, t)=P_{r,a_{1}}(x, t)+P_{r,d_{1}}(x, t)$ where, from equations~(\ref{Pra2}) and ~(\ref{eq:full_PDF}) for $n=1$,
\begin{subequations} 
\begin{align}
P_{r,a_{1}}(x, t) &=P_{r,1}^{*}(x)-e^{-r t} P_{r,1}^{*}(x) * G_{0}(x, t)   \\
P_{r,d_{1}}(x, t) &=e^{-r t} G_{0}(x, t) .
\end{align}
\end{subequations}
Our goal is to characterise the approach to steady-state of $P_{r,a_{1}}$. First, we define the fractional deviation from the NESS according to
\begin{equation}
\label{Z1}
Z_{r,a_{1}}(x, t)=1-\frac{P_{r,a_{1}}(x, t)}{P_{r,1}^{*}(x)}=\frac{e^{-r t} P_{r,1}^{*}(x) * G_{0}(x, t)}{P_{r,1}^{*}(x)} . 
\end{equation}
Note that $Z_{r,a_1}(x,t)$ is non-negative (since $P_{r,a_1}(x,t) \leq P_{r,1}^*(x)$ for all $t\geq 0$~\cite{Bressloff21}), decreases monotonically from $Z_{r,a_1}(x,0)=1$ to $Z_{r,a_1}(x,\infty)=0$, and can therefore be interpreted as a survival probability for the local establishment of the NESS at position $x$. Interpreting $Z_{r,a_{1}}$ as a survival probability, as justified above, we define the accumulation time \cite{Bressloff21,Berez10,Berez11,Gordon11}, the effective time taken to relax to the NESS up to point $x$, analogously to the mean first-passage time of $Z_{r,a_{1}}$
\begin{equation}
\label{eq:accum_def}
T_{r,a_{1}}(x) = -\int_{0}^{\infty}\mathrm{d}t\; t\,\partial_t Z_{r,a_{1}}(x,t) = \int_{0}^{\infty}\mathrm{d}t\; Z_{r,a_{1}}(x,t).
\end{equation}
The equality between the two expressions
in~(\ref{eq:accum_def}) follows from integration by parts:
\begin{equation}
-\int_0^\infty \mathrm{d}t\, t\,\partial_t Z
= -\Big[t\,Z\Big]_0^\infty + \int_0^\infty\mathrm{d}t\, Z
= \int_0^\infty \mathrm{d}t\, Z,
\end{equation}
where the boundary term vanishes provided $t\,Z_{r,a_1}(x,t)\to 0$ as
$t\to\infty$. This condition holds here because $Z_{r,a_1}(x,t)$ decays
exponentially in $t$ for fixed $x$ (as can be seen from the explicit
form~(\ref{Z1}) and the exponential decay of $e^{-rt}$), so
$t\,Z_{r,a_1}(x,t) = \mathcal{O}(t\,e^{-rt})\to 0$. We verify below that
the same condition holds for the nested case. Performing the integral over time and using identity~(\ref{eq:laplace_G0}), we arrive at
\begin{equation}
\label{T1}
T_{r,a_{1}}(x)=\frac{P_{r,1}^{*}(x) * \int_{0}^{\infty} d t e^{-r t} G_{0}(x, t)}{P_{r,1}^{*}(x)}=\frac{P_{r,1}^{*}(x) * P_{r,1}^{*}(x)}{r P_{r,1}^{*}(x)} 
\end{equation}

It remains to evaluate the convolution
$P_{r,1}^*(x)*P_{r,1}^*(x)$. Since $P_{r,1}^*(x) = \frac{\alpha}{2}
e^{-\alpha|x|}$ is a Laplace distribution, its convolution with itself is
a known closed form. In Fourier space,
\begin{equation}
\widetilde{P_{r,1}^* * P_{r,1}^*}(k)
= \left[\widetilde{P}_{r,1}^*(k)\right]^2
= \frac{r^2}{(r+Dk^2)^2}
= \widetilde{P}_{r,2}^*(k),
\end{equation}
so $P_{r,1}^* * P_{r,1}^* = P_{r,2}^*(x)$, the two-particle steady-state
distribution. In real space, using equation~(\ref{ssn}) with $n=2$
and the identity $K_{3/2}(z) = \sqrt{\pi/(2z)}\,e^{-z}(1+1/z)$~\cite{Abramowitz72},
\begin{equation}
\label{eq:conv_laplace}
P_{r,1}^*(x)*P_{r,1}^*(x)
= P_{r,2}^*(x)
= \frac{\alpha}{2}\,e^{-\alpha|x|}\left(\frac{\alpha|x|+1}{2}\right)
  \cdot \frac{2}{\alpha|x|+1} \cdot \frac{\alpha|x|+1}{2}
= \frac{\alpha}{4}\,e^{-\alpha|x|}(\alpha|x|+1).
\end{equation}
Substituting into~(\ref{T1}), we find that $T_{r,a_{1}}$ can be written as
\begin{equation}
T_{r,a_{1}}(x)=\frac{\frac{\alpha}{2} e^{-\alpha|x|}\left(\frac{\alpha|x|+1}{2}\right)}{r \frac{\alpha}{2} e^{-\alpha|x|}}=\frac{|x|}{\sqrt{4 r D}}+\frac{1}{2 r} 
\end{equation}

We can proceed similarly with $P_{r,d_{1}}(x,t)$ to find
\begin{equation}
T_{r,d_{1}}(x)=-\frac{1}{r} ,
\end{equation}
which is the `negative time' associated with the decay of the initial condition of the particle. For the remainder of the paper we will not be concerned with this quantity; as discussed in \cite{Bressloff21}, $P_{r,d_{1}}$ overshoots $P^*_{r,1}$ and hence $Z_{r,d_1}$ is not everywhere non-negative, so we cannot interpret it as a survival probability or define a meaningful accumulation time for it. Furthermore, at large $|x|$ the approach to the NESS is encoded entirely in $P_{r,a_1}$, since $P_{r,d_1}(x,t) = e^{-rt}G_0(x,t)$ decays uniformly in $x$.

Interestingly, using the definition of the length scale $\alpha^{-1}$, we can rewrite $T_{r,a_{1}}(x)$ as
\begin{equation}
\label{T1a}
T_{r,a_{1}}(x)=\frac{|x|+\alpha^{-1}}{\sqrt{4 D r}} .
\end{equation}

\begin{figure}[t!]
\begin{center}
\includegraphics[width=\linewidth]{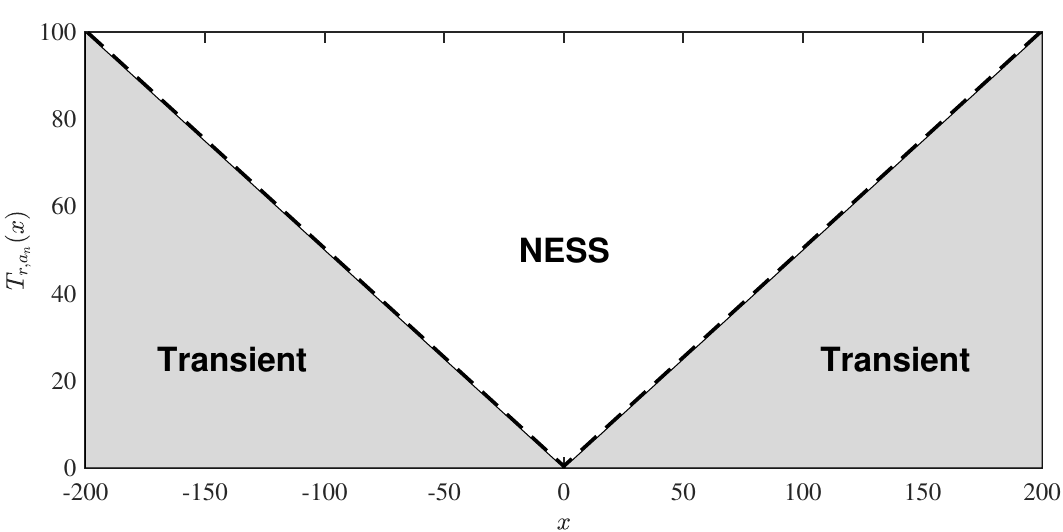}
\caption{\textit{Single particle accumulation time.} The accumulation time of a single stochastic resetter $T_{r,a_1}(x)$, equation (\ref{T1a}), for $r=1$ and $D=1$. The curve separates space into a NESS region, where the particle has reached its non-equilibrium steady state, and a transient region, where the particle is yet to reach steady state.}\label{fig3}
\end{center}
\end{figure}

\noindent The accumulation time grows linearly with $|x|$, with slope $1/\sqrt{4Dr}$ and intercept $\alpha^{-1}/\sqrt{4Dr}$. This can be interpreted as follows: the NESS propagates outward from the origin with a characteristic \emph{effective speed} $v^* = \sqrt{4Dr}$, but with an inherent delay $\alpha^{-1}/\sqrt{4Dr} = 1/(2r)$ that reflects the time required for the initial delta function to spread over the typical reset length $\alpha^{-1}$ before the resetting mechanism can establish the NESS. We stress that $v^* = \sqrt{4Dr}$ is not a sharp wavefront velocity in the sense of a discontinuity in the distribution; rather, it characterises the \emph{effective} rate at which the accumulation time grows with distance, and is closely related to the ballistic propagation of the NESS boundary identified in Ref.~\cite{Majumdar15}. Interpreting $\sqrt{4 D r}$ as a velocity scale, then the second term in the accumulation time measures how long on average the particle takes to initially explore space. The accumulation time~(\ref{T1a}) separates space into a NESS region (where the particle has effectively reached its nonequilibrium steady state) and a transient region (where it has not), as illustrated in Figure~\ref{fig3}.

\subsection{Accumulation times for two particles with equal rates}
We now calculate the accumulation time for the second particle in a nested stochastic resetting chain with equal rates. The decomposed probability distribution is $P_{r,2}(x, t)= P_{r,a_{2}}(x, t)+P_{r,d_{2}}(x, t)$ with
\begin{subequations}
\begin{align}
P_{r,a_{2}} & =P_{r,2}^{*}(x)-e^{-r t}\left[P_{r,2}^{*}(x)+r t P_{r,1}^{*}(x)\right] * G_{0}(x, t)  ,\\
P_{r,d_{2}} & =e^{-r t}[r t+1] G_{0}(x, t) .
\end{align}
\end{subequations}
These follow directly from equation~(\ref{Pra2}) with $n=2$ and equation~(\ref{Prd}) with $n=2$, respectively. The fractional deviation from the NESS for particle $2$, $Z_{r,a_{2}}$, is defined analogously to equation (\ref{Z1}) and reads
\begin{equation}
\label{eq:Z2}
Z_{r,a_{2}}=1-\frac{P_{r,a_{2}}(x, t)}{P_{r,2}^{*}(x)}=\frac{e^{-r t}\left[P_{r,2}^{*}(x)+r t P_{r,1}^{*}(x)\right] * G_{0}(x, t)}{P_{r,2}^{*}(x)} .
\end{equation}
As in the single-particle case, the boundary term $t\,Z_{r,a_2}(x,t)\to 0$ as $t\to\infty$ holds because the numerator of~(\ref{eq:Z2}) decays as $\mathcal{O}(t\,e^{-rt})$, so the accumulation time is well-defined. The accumulation time of particle 2 is then
\begin{eqnarray}
\label{eq:T2_exact}
T_{r,a_{2}}(x) & =\int_{0}^{\infty} d t Z_{r,a_{2}}(x, t)  \nonumber \\
& =\frac{\left[P_{2}^{*}(x) * \int_{0}^{\infty} d t e^{-r t} G_{0}(x, t)\right]+\left[P_{r,1}^{*}(x) * \int_{0}^{\infty} d t r t e^{-r t} G_{0}(x, t)\right]}{r P_{r,2}^{*}(x)}\nonumber \\
& =\frac{2\left[P_{r,2}^{*}(x) * P_{r,1}^{*}(x)\right]}{r P_{r,2}^{*}(x)}  \nonumber\\
& =\frac{3+3 \alpha|x|+\alpha^{2}|x|^{2}}{2 r[1+\alpha|x|]} .
\end{eqnarray}
For $|x| \gg \alpha^{-1}$, we find
\begin{equation}
\label{eq:T2_large_x}
T_{r,a_{2}}(x) \sim \frac{|x|+2 \alpha^{-1}}{\sqrt{4 D r}} ,
\end{equation}
which has the same linear structure as the single-particle result~(\ref{T1a}), with the delay increased from $\alpha^{-1}$ to $2\alpha^{-1}$. This linear increase from particle~$1$ to
particle~$2$ can be understood as follows: particle~$1$ must first spread over a length $\alpha^{-1}$ before it can establish its NESS, contributing a delay of $\alpha^{-1}/\sqrt{4Dr}$. Particle~$2$ then resets to the instantaneous position of particle~$1$, and must itself spread over an additional length $\alpha^{-1}$ on average, contributing an equal additional delay. The total delay is therefore $2\alpha^{-1}/\sqrt{4Dr}$, consistent with~(\ref{eq:T2_large_x}).

\subsection{\texorpdfstring{Accumulation times for $N$-particles nested with equal rates}{Accumulation times for N-particles nested with equal rates}}
We can define the fractional deviation from the NESS for general $n$ according to
\begin{equation}
\label{Zn}
Z_{r,a_{n}}(x, t)=1-\frac{P_{r,a_{n}}(x, t)}{P_{r,n}^{*}(x)}=\frac{\sum_{j=0}^{n-1} \frac{(r t)^{j}}{j!} e^{-r t} P_{r,n-j}^{*}(x) * G_{0}(x, t)}{P_{r,n}^{*}(x)} .
\end{equation}
which follows directly from equation~(\ref{Pra2}). The boundary condition $t\,Z_{r,a_n}(x,t)\to 0$ as $t\to\infty$ holds for all $n\geq 1$, since the numerator of~(\ref{Zn}) decays as $\mathcal{O}(t^{n-1} e^{-rt})$ for fixed $x$, so the accumulation time is well-defined for all particles in the chain. The accumulation time of particle $n$ is
\begin{eqnarray}
T_{r,a_{n}}(x) & =\int_{0}^{\infty} d t Z_{r,a_{n}}(x, t)  \nonumber \\
& =\frac{\sum_{j=1}^{n} P_{r,n-j+1}^{*}(x) * r \int_{0}^{\infty} d t \frac{(r t)^{j-1}}{(j-1)!} e^{-r t} G_{0}(x, t)}{r P_{r,n}^{*}(x)}  \nonumber\\
& =\frac{\sum_{j=1}^{n} P_{r,n-j+1}^{*}(x) * P_{r,j}^{*}(x)}{r P_{r,n}^{*}(x)}  \nonumber\\
& =\frac{n\left(P_{r,n}^{*}(x) * P_{r,1}^{*}(x)\right)}{r P_{r,n}^{*}(x)},\label{tran}
\end{eqnarray}
where the second line uses the generalised identity
\begin{equation}
\label{eq:laplace_G0_general}
r\int_0^\infty \frac{(rt)^{j-1}}{(j-1)!}\,e^{-rt}\,G_0(x,t)\,\mathrm{d}t = P_{r,j}^*(x),
\end{equation}
which follows from equation~(\ref{Pra}) by taking $t\to\infty$ and recognising that the left-hand side is precisely the steady-state accumulation integral for particle $j$, equal to $P_{r,j}^*(x)$ by equation~(\ref{eq:limits_a}). The claim that 
\begin{equation}
\label{eq:sum_simplification}
\sum_{j=1}^{n} P_{r,n-j+1}^{*}(x)*P_{r,j}^{*}(x)
= n\left(P_{r,n}^{*}(x)*P_{r,1}^{*}(x)\right)
\end{equation}
is non-trivial and we justify it now. Since $P_{r,n}^*(x)$ is the $n$-fold convolution of $P_{r,1}^*(x)$ with itself (as established in Section~\ref{subsec:FP}), we have in Fourier space
\begin{equation}
\widetilde{P}_{r,m}^*(k) = \left[\widetilde{P}_{r,1}^*(k)\right]^m = \left(\frac{r}{r+Dk^2}\right)^m.
\end{equation}
Therefore, in Fourier space, the left-hand side of~(\ref{eq:sum_simplification}) becomes
\begin{subequations}
\begin{align}
\sum_{j=1}^{n}\widetilde{P}_{r,n-j+1}^*(k)\cdot\widetilde{P}_{r,j}^*(k)
&= \sum_{j=1}^{n}  \left(\frac{r}{r+Dk^2}\right)^{n-j+1}  \left(\frac{r}{r+Dk^2}\right)^{j} \\
& = \sum_{j=1}^{n}  \left(\frac{r}{r+Dk^2}\right)^{n+1} 
= n\,\widetilde{P}_{r,n+1}^*(k),
\end{align}
\end{subequations}
which in real space gives $n\,P_{r,n+1}^*(x) = n\,P_{r,n}^*(x)*P_{r,1}^*(x)$, confirming~(\ref{eq:sum_simplification}). From~(\ref{tran}), we can recover the single-particle result~(\ref{T1}) by setting $n=1$. We can also verify the two-particle result: setting $n=2$ gives $T_{r,a_2}(x) = 2(P_{r,2}^**P_{r,1}^*)/(rP_{r,2}^*)$, consistent with the third line of~(\ref{eq:T2_exact}).

To study the large $|x|$ behaviour of~(\ref{tran}), we use the fact that $P_{r,n}^*(x)*P_{r,1}^*(x) = P_{r,n+1}^*(x)$ (as established above), so
\begin{equation}
\label{eq:Tan_ratio}
T_{r,a_n}(x) = \frac{n\,P_{r,n+1}^*(x)}{r\,P_{r,n}^*(x)} = \frac{(\alpha|x|/2)\,K_{n+1/2}(\alpha|x|)}{\alpha\,K_{n-1/2}(\alpha|x|)}.
\end{equation}
For the large $|x|$ expansion, we use the asymptotic expansion of the modified Bessel function of the second kind for large
argument~\cite{Abramowitz72},
\begin{equation}
\label{eq:Knu_asymp}
K_\nu(z) \sim \sqrt{\frac{\pi}{2z}}\,e^{-z} \sum_{k=0}^{m}\frac{(\nu,k)}{(2z)^k} + \mathcal{O}(z^{-m-1}), \quad z\to\infty,
\end{equation}
where $(\nu,k) = \frac{\Gamma(\nu+k+1/2)}{k!\,\Gamma(\nu-k+1/2)}$ are the Hankel coefficients. To leading and subleading order, $K_\nu(z) \sim \sqrt{\pi/(2z)}\,e^{-z}[1 + (4\nu^2-1)/(8z) +
\mathcal{O}(z^{-2})]$. Taking the ratio:
\begin{equation}
\label{aBessel}
\frac{K_{n+1/2}(\alpha|x|)}{K_{n-1/2}(\alpha|x|)}
= 1 + \frac{n}{\alpha|x|} + \mathcal{O}\!\left(1/x^{2}\right),
\quad |x|\gg(n-1)\alpha^{-1},
\end{equation}
where the subleading correction $n/(\alpha|x|)$ follows from the difference of the Hankel coefficients at orders $\nu = n+1/2$ and
$\nu = n-1/2$. Substituting~(\ref{aBessel}) into~(\ref{eq:Tan_ratio}) and retaining terms to zeroth order in $1/(\alpha|x|)$, i.e.\ working to the same order as in equation~(\ref{T1a}), we show that the accumulation time of particle $n$ in a
nested resetting chain with equal rates $r$ and diffusivity $D$ satisfies
\begin{equation}
\label{tan}
T_{r,a_{n}}(x) \sim \frac{|x|+n\alpha^{-1}}{\sqrt{4Dr}}, \qquad |x|\gg n\alpha^{-1}
\end{equation}
Once again, the delay $n\alpha^{-1}/\sqrt{4Dr}$ increases \emph{linearly} with the particle index $n$, with the same effective speed $\sqrt{4Dr}$ as the single-particle case~(\ref{T1a}).

\begin{figure}[t!]
\begin{center}
\includegraphics[width=\linewidth]{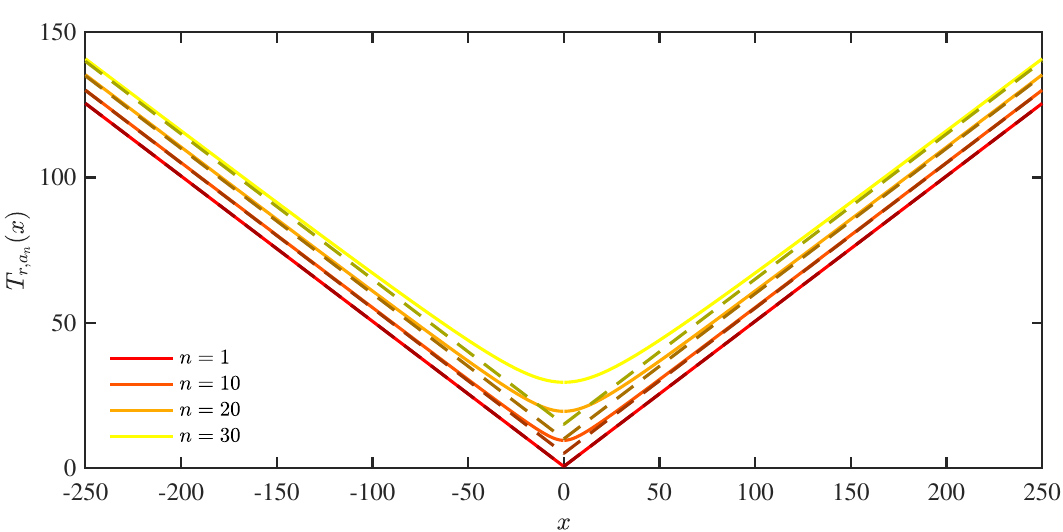}
\caption{\textit{Accumulation time of nested resetters with equal rates.} The accumulation time $T_{r,a_n}(x)$ of particle $n$ in a nested resetting chain with equal rates, equation (\ref{tran}), and the corresponding large $|x|$ approximation, equation (\ref{tan}), plotted for a range of $n$ with $r=1$ and $D=1$. Solid lines indicate the exact results; dashed lines indicate the large $|x|$ approximation.}
\label{fig:equalaccum}
\end{center}
\end{figure}

We compare the exact result~(\ref{tran}) and the large $|x|$ approximation~(\ref{tan}) in Figure~\ref{fig:equalaccum}. The linear increase in the delay with $n$ is clearly visible in both the exact and asymptotic results. Here again, for the particle to reach its NESS, information must propagate sequentially down the nested resetting chain. Since all particles share the same length scale $\alpha^{-1}$ (a consequence of equal diffusivity and equal resetting rate), these delays accumulate linearly, yielding a total delay of $n\alpha^{-1}/\sqrt{4Dr}$ for particle $n$.

\section{Nested Resetting with Heterogeneous Rates}
\label{sec:hetero}

We now extend our results to systems where particles can reset with heterogeneous rates. We first assume non-degeneracy, such that all particles reset with distinct rates, but then relax this assumption and allow for more than one particle to share the same resetting rate. The key structural observation here is that, in the non-degenerate case, the time-dependent distribution can be written as a linear combination of single-particle distributions via a partial-fraction decomposition of the renewal equation. In the degenerate case, this decomposition generalises to include higher-order poles, and is most naturally handled via the Fokker-Planck equation.

\subsection{Renewal approach for non-degenerate resetting rates}
Using renewal theory, one can write a last renewal equation for the probability density function of an $n^{\text {th }}$ nested resetter in terms of the probability density of the $n-1^{\text {th }}$ particle in the chain, as in equation (\ref{renn}):
\begin{equation}
\label{Pdiff}
P_{\mathbf{r},n}(x,t) = e^{-r_n t}\,G_{0}(x,t) + r_n\int_{0}^{t}\mathrm{d}\tau\, e^{-r_n\tau} \int_{-\infty}^{+\infty}\mathrm{d}\xi\; G_{0}(x-\xi,\tau)\,P_{\mathbf{r},n-1}(\xi,t-\tau),
\end{equation}
where $\r=(r_1,\ldots,r_N)$.
Particle $n$ now resets with rate $r_{n}$, and typically diffuses a distance $\alpha_{n}^{-1}=\sqrt{D / r_{n}}$ between resets. 

As in the equal-rate case (equation~(\ref{renn2})), one can eliminate the explicit dependence on the hierarchy of distributions by iterating~(\ref{Pdiff}). For non-degenerate rates, the result takes the form of a \emph{partial-fraction decomposition} in the Laplace variable: the Laplace transform of $P_{\mathbf{r},n}(x,t)$ with respect to $t$ is a rational function of $s$ with simple poles at $s = -r_j$, $j=1,\ldots,n$, and the partial-fraction decomposition of this rational function yields the following representation in the time domain:
\begin{align}
P_{\r,n}(x, t) & =\sum_{j=1}^{n}\left[e^{-r_{j} t} G_{0}(x, t)+r_{j} \int_{0}^{t} \mathrm{d} \tau e^{-r_{j} \tau} G_{0}(x, \tau)\right] \prod_{i=1, i \neq j}^{n} \frac{r_{i}}{r_{i}-r_{j}} \nonumber \\
& =\sum_{j=1}^{n} P_{r_j,1}\left(x, t\right) \prod_{i=1, i \neq j}^{n} \frac{r_{i}}{r_{i}-r_{j}} ,
\label{Pnhet}
\end{align}
where $P_{r_j,1}\left(x, t \right)$ is the single particle PDF with resetting rate $r_{j}$ and the weights $\prod_{i \neq j} r_i/(r_i-r_j)$ are the Lagrange basis coefficients of the partial-fraction decomposition. The result states that the PDF of the $n$-th particle is a linear combination of single-particle PDFs, one for each resetting rate in the chain. This representation is valid provided all rates are distinct ($r_i\neq r_j$ for $i\neq j$); the degenerate case is treated in Section~\ref{sec:deg}.

Note that the result~(\ref{Pnhet}) can be established by induction on $n$, using the same approach as in Section~\ref{subsec:renewal}. The base case $n=1$ is equation~(\ref{ren1}). Assuming~(\ref{Pnhet}) holds for $n-1$, one substitutes into~(\ref{Pdiff}) and applies the Chapman--Kolmogorov identity~(\ref{eq:CK}), then uses the partial-fraction identity
\begin{equation}
\label{eq:pf_het}
\frac{r_n}{s+r_n}\cdot\prod_{j=1}^{n-1}\frac{r_j}{s+r_j} = \sum_{j=1}^{n}\frac{r_j}{s+r_j}\prod_{i\neq j}^{n}\frac{r_i}{r_i-r_j}
\end{equation}
to collapse the result into the form~(\ref{Pnhet}) for $n$.

\begin{figure}[t!]
\begin{center}
\includegraphics[width=\linewidth]{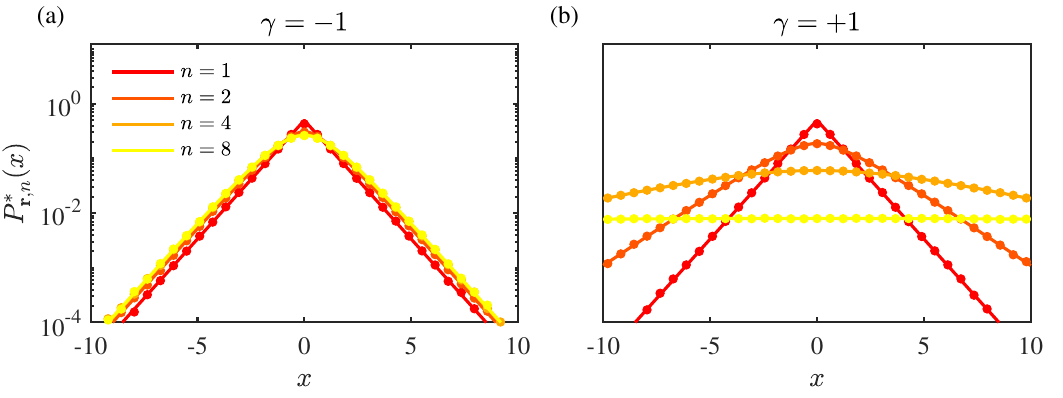}
\caption{\textit{Steady-state distributions $P^*_{\mathbf{r},n}(x)$ of nested resetters with geometric rates.} Steady-state distributions of nested resetters with (a) geometrically increasing resetting rates $(r_i = e^{i-1})$ and (b) geometrically decreasing resetting rates $(r_i = e^{1-i})$, equation (\ref{ssnondeg}), both with diffusivity $D=1$ for $n\in\{1,2,4,8\}$. Solid lines indicate analytical results and symbols indicate numerical simulations.}
\label{fig:geometricss}
\end{center}
\end{figure}

Further, we can perform a decomposition analogous to equations (\ref{Pa}) and (\ref{Pd}):
\begin{subequations}
\begin{align}
\label{PraND}
& P_{\r,a_{n}(x, t)}=\sum_{j=1}^{n} P_{r_j,a_{1}}\left(x, t \right) \prod_{i=1, i \neq j}^{n} \frac{r_{i}}{r_{i}-r_{j}}  ,\\
& P_{\r,d_{n}(x, t)}=\sum_{j=1}^{n} P_{r_j,d_{1}}\left(x, t \right) \prod_{i=1, i \neq j}^{n} \frac{r_{i}}{r_{i}-r_{j}} .
\label{PrdND}
\end{align}
\end{subequations}
Here $P_{r_j,a_1}(x,t)$ and $P_{r_j,d_1}(x,t)$ are the accumulating and decaying components of a single-particle distribution with resetting rate $r_j$, as defined in equations~(\ref{Pra}) and~(\ref{Prd}). We can easily take the long-time limit $t \rightarrow \infty$ of equations (\ref{Pnhet}) to give
\begin{equation}
\label{ssnondeg}
P_{\r,n}^{*}(x)=\sum_{j=1}^{n} P_{r_j,1}^{*}\left(x \right) \prod_{i=1, i \neq j}^{n} \frac{r_{i}}{r_{i}-r_{j}},
\end{equation}
as numerically confirmed in Figure \ref{fig:geometricss} for geometric resetting rates, where particle $i$ resets with rate $r_i = r\,e^{-\gamma(i-1)}$. Moving along the chain, we see either a monotonically increasing, unbounded resetting rate for $\gamma<0$, a monotonically decreasing resetting rate that goes to 0 for $\gamma>0$, or constant resetting rate $r$ for $\gamma=0$.

As expected, both the steady-state and full time-dependent distributions are non-singular in the limit $r_i\to r_j$ for all $i,j$, since the Lagrange weights $\prod_{i\neq j}r_i/(r_i-r_j)$ have removable singularities at $r_i=r_j$, as we now verify.
For example, the steady state distribution can be rewritten in the form
\begin{equation}
\label{ssnondeg2}
P_{\r,n}^{*}(x)=(-1)^{n-1}\left(\prod_{i=1}^{n} r_{i}\right) \sum_{j=1}^{n} \frac{P_{r_j,1}^{*}\left(x  \right)}{r_{j} \prod_{i \neq j}\left(r_{j}-r_{i}\right)}  
\end{equation}
Taking the limit $r_j\rightarrow r$ using L'H\^opital's rule yields
\begin{equation*}
\left (\prod_{j=1}^n\lim_{r_j\rightarrow r}\right )P_{\r,n}^{*}(x)=\frac{(-1)^{n-1} r^{n}}{(n-1)!} \frac{\mathrm{d}^{n-1}}{\mathrm{d} r^{n-1}}\left(\frac{P_{r,1}^{*}(x )}{r}\right) 
\end{equation*}
which, using the modified Bessel function identities~\cite{Abramowitz72}
\begin{equation}
\frac{\mathrm{d}}{\mathrm{d} z}\left(z^{-\nu} K_{\nu}(z)\right)=-z^{-\nu} K_{\nu+1}(z)  
\end{equation}
and the iterated form
\begin{equation}
\left(\frac{1}{z} \frac{\mathrm{d}}{\mathrm{d} z}\right)^{m}\left(z^{-\nu} K_{\nu}(z)\right)=(-1)^{m} z^{-\nu-m} K_{\nu+m}(z) 
\end{equation}
indeed reproduces the equal rates steady state distribution (\ref{ssn}). This confirms that the non-degenerate formula~(\ref{ssnondeg}) has a well-defined limit as any subset of rates become equal, and that this limit coincides with the degenerate formula derived below.

So far we have assumed that the rates are non-degenerate. However, to complete a more general picture of nested stochastic resetting we need to allow for partially degenerate rates (i.e.\ any pair of particles can share the same resetting rate). In this case, the inductive step that allows us to write the renewal equation as a linear combination of single-particle distributions involves higher-order poles in the Laplace variable, and is far less transparent. We therefore turn to the Fokker-Planck equation, which handles degeneracy naturally through the partial-fraction decomposition of rational functions with repeated poles.

\subsection{Fokker-Planck approach for degeneracy}
\label{sec:deg}

Consider a chain of $n$ particles such that $m_i$ particles (that are not necessarily neighbors in the resetting hierarchy) reset at the same rate $r_i$, subject to $\sum_{i=1}^{M}m_i=n$, where $M\leq n$ is the number of distinct resetting rates. For simplicity, we assume there exists $M\leq n$ unique resetting rates in the chain of length $n$. Let $P_{M,j}(x,t)$  denote the PDF of the $j$-th particle in the chain for a given degeneracy, which depends on $\r=(r_1,\ldots,r_M)$ and ${\bf m}=(m_1,\ldots,m_M)$. Rather than working from the last renewal equation, we start from the Fokker-Planck equation,
\begin{equation}
\label{eq:FP_deg}
\partial_{t} P_{M,j}(x, t)=D \nabla^{2} P_{M,j}(x, t)-\mathcal{R}_{j} P_{M,j}(x, t)+\mathcal{R}_{j} P_{M,j-1}(x, t), \quad P_{M,j}(x, 0)=\delta(x) 
\end{equation}
where particle $j$ resets with rate $\mathcal{R}_j \in \mathbf{r}$ for $j=1,\ldots,n$. The steady-state distribution of particle $n$ in Fourier space follows
immediately from iterating the steady-state recursion:
\begin{equation}
\label{sso}
\widetilde{P}_{M,n}^{*}(k)=\prod_{i=1}^{M}\left(\frac{r_{i}}{r_{i}+D k^{2}}\right)^{m_{i}}  .
\end{equation}
This is a product of $M$ factors, each of which is the characteristic function of an $m_i$-particle equal-rate chain with rate $r_i$ (see\ equation~(\ref{PPn})). The partial-fraction decomposition of this expression in the variable $\mathcal{X}=Dk^2$ yields the real-space steady-state distribution.

We can write the general partial fraction decomposition of any expression of the form (\ref{sso}) as
\begin{equation}
\prod_{i=1}^{M}\left(\frac{r_{i}}{r_{i}+\mathcal{X}}\right)^{m_{i}}=\sum_{i=1}^{M} \sum_{j=1}^{m_{i}} A_{i, j}\left(\frac{1}{r_{i}+\mathcal{X}}\right)^{j} ,
\label{eq:pf_deg}
\end{equation}
where $\mathcal{X}=Dk^2$ is a non-negative real variable and
the coefficients $A_{i,j}$ are determined by the Laurent expansion of the
left-hand side around each pole $\mathcal{X}=-r_i$. The coefficients $A_{i,j}$ are thus uniquely defined by
\begin{eqnarray}
\fl A_{i,j}
&=& \frac{r_i^{m_i}}{(m_i-j)!}
    \Biggl[\frac{\mathrm{d}^{m_i-j}}{\mathrm{d}\mathcal{X}^{m_i-j}}
    \prod_{\ell\neq i}^{M}
    \left(\frac{r_\ell}{r_\ell+\mathcal{X}}\right)^{m_\ell}\Biggr]_{\mathcal{X}=-r_i} \nonumber\\
\fl &=& \frac{r_i^{m_i}}{(m_i-j)!}
    \sum_{\substack{\sum_a k_a=m_i-j \\ k_i=0}}
    \binom{m_i-j}{k_1,k_2,\ldots,k_n}
    \prod_{\ell\neq i}^{M}
    \Biggl[\frac{\mathrm{d}^{k_\ell}}{\mathrm{d}\mathcal{X}^{k_\ell}}
    \left(\frac{r_\ell}{r_\ell+\mathcal{X}}\right)^{m_\ell}
    \Biggr]_{\mathcal{X}=-r_i} \label{eq:Aij}\\
\fl &=& \frac{(-1)^{m_i-j}r_i^{m_i}}{(m_i-j)!}
    \sum_{\substack{\sum_a k_a=m_i-j \\ k_i=0}}
    \binom{m_i-j}{k_1,k_2,\ldots,k_n}
    \prod_{\ell\neq i}^{M}
    \frac{(n_\ell+k_\ell-1)!}{(n_\ell-1)!}
    \frac{r_\ell^{n_\ell}}{(r_\ell-r_i)^{m_\ell+k_\ell}}. \nonumber
\end{eqnarray}

To illustrate the formula~(\ref{eq:Aij}), consider $M=2$ distinct rates $r_1\neq r_2$, each with multiplicity $m_1=m_2=1$ (so $n=2$). The decomposition~(\ref{eq:pf_deg}) reads
\begin{equation}
\frac{r_1}{r_1+\mathcal{X}}\cdot\frac{r_2}{r_2+\mathcal{X}}= \frac{A_{1,1}}{r_1+\mathcal{X}} + \frac{A_{2,1}}{r_2+\mathcal{X}}.
\end{equation}
Since $m_i=1$ and $j=1$, the formula~(\ref{eq:Aij}) gives $A_{i,j}$ with $m_i-j=0$, so the sum over $k_a$ contains only the term $k_a=0$ for all $a$:
\begin{equation}
A_{1,1} = r_1 \cdot \frac{r_2}{r_2-r_1} = \frac{r_1 r_2}{r_2-r_1},
\qquad
A_{2,1} = r_2 \cdot \frac{r_1}{r_1-r_2} = \frac{r_1 r_2}{r_1-r_2}.
\end{equation}
One can verify directly that $A_{1,1}/(r_1+\mathcal{X}) + A_{2,1}/(r_2+\mathcal{X}) = r_1 r_2/[(r_1+\mathcal{X})(r_2+\mathcal{X})]$, which equals the left-hand side. The steady-state distribution~(\ref{dec}) then gives
\begin{equation}
P_{2,2}^*(x) = \frac{A_{1,1}}{r_1}P_{r_1,1}^*(x) + \frac{A_{2,1}}{r_2}P_{r_2,1}^*(x) = \frac{r_2}{r_2-r_1}P_{r_1,1}^*(x) + \frac{r_1}{r_1-r_2}P_{r_2,1}^*(x),
\end{equation}
recovering the non-degenerate result~(\ref{ssnondeg}) with $n=2$.

\begin{figure}[t!]
\begin{center}
\includegraphics[width=\linewidth]{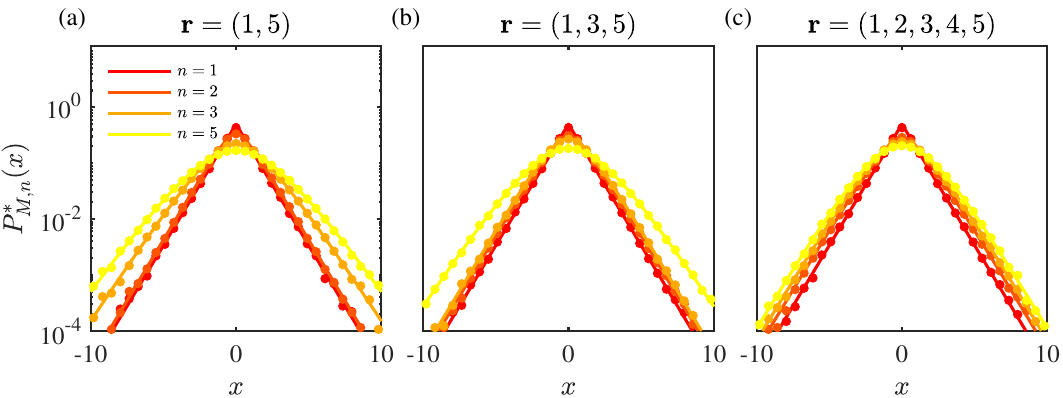}
\caption{\textit{Steady-state distribution $P^*_{M,n}(x)$ of nested resetters with periodic rates.} Steady-state distributions of nested resetters with periodic resetting rates, equation (\ref{dec}), and diffusivity $D=1$ for $n\in\{1,2,3,5\}$, where $n$ has been chosen to reflect the sizes of the respective periods: (a) $\mathbf{r} = (1,5)$; (b) $\mathbf{r} = (1,3,5)$; (c) $\mathbf{r} = (1,2,3,4,5)$.  Solid lines indicate analytical results and symbols indicate numerical simulations.}
\label{fig:periodicss}
\end{center}
\end{figure}

Importantly, this allows us to write the steady-state distribution of particle $n$ in real space as
\begin{equation}
\label{dec}
P_{M,n}^{*}(x)=\sum_{i=1}^{M} \sum_{j=1}^{m_{i}} \frac{A_{i, j}}{r_{i}^{j}} P_{r_i,j}^{*}\left(x \right) ,
\end{equation}
as numerically confirmed in Figure \ref{fig:periodicss} for periodic resetting rates, where particle $i$ resets with rate $r_j$ with $j\equiv i\bmod M$. The steady-state distribution is thus a weighted sum of equal-rate $j$-particle distributions, where the weights $A_{i,j}/r_i^j$ are determined by the partial-fraction decomposition~(\ref{eq:pf_deg}).

We immediately see from the steady-state distribution~(\ref{dec}) that for periodic rates the characteristic function factorises into blocks of equal resetting rate:
\begin{equation}
\widetilde{P}_{M,n}^{*}(k)=\prod_{i=1}^{M}\left(\frac{r_{i}}{r_{i}+D k^{2}}\right)^{m} \times \prod_{j=1}^{n-m M} \frac{r_{j}}{r_{j}+D k^{2}},  
\end{equation}
where $m = \lfloor n/M \rfloor$ is the number of complete periods and the second product runs over the $n-mM$ residual particles that make up the incomplete final period.

In the limit $r_{i} \rightarrow r$ for all $i$, such that all particles reset with the same rate $r$, we find 
\begin{equation}
\left (\prod_{j=1}^M\lim_{r_j\rightarrow r}\right )P_{M,n}^{*}(x)=\sum_{j=1}^{n} \frac{A_{1, j}}{r^{j}} P_{r,j}^{*}(x )  
\end{equation}
with
\begin{equation}
A_{1, j}= \begin{cases}r^{n} & j=n,   \\ 0 & \mathrm{otherwise},\end{cases}
\end{equation}
which follows from the fact that in this limit the partial-fraction decomposition~(\ref{eq:pf_deg}) collapses to a single $n$-th order pole at $\mathcal{X}=-r$. Hence, we recover the equal rates steady-state distribution (\ref{ssn}). 

On the other hand, in the limit $n_{i} \rightarrow 1$ for all $i=1,\ldots,M$ with $M=n$, we find
\begin{equation}
\left (\prod_{j=1}^M\lim_{n_j\rightarrow 1}\right )P_{M,n}^{*}(x)=\sum_{i=1}^{n} \frac{A_{i, 1}}{r_{i}} P_{r_i,1}^{*}\left(x  \right),
\end{equation}
with
\begin{equation}
A_{i, 1}=r_{i} \prod_{\ell \neq i}^{n} \frac{r_{\ell}}{r_{\ell}-r_{i}}  
\end{equation}
recovering the non-degenerate result (\ref{ssnondeg}). 

If, once again, we decompose $P_{M,n}(x, t)=P_{M,a_{n}}(x, t)+P_{M,d_{n}}(x, t)$, as in equations (\ref{Pa}) and (\ref{Pd}) with
\begin{subequations}
\begin{eqnarray}
\fl & \partial_{t} P_{M,a_{n}} = D \partial^{2}_x P_{M,a_{n}}-\mathcal{R}_{n} P_{M,a_{n}}+\mathcal{R}_{n} P_{M,a_{n-1}}, \quad P_{M,a_{n}}(x, 0)=0  \\
\fl & \partial_{t} P_{M,d_{n}} = D \partial^{2}_x P_{M,d_{n}}-\mathcal{R}_{n} P_{M,d_{n}}+\mathcal{R}_{n} P_{M,d_{n-1}}, \quad P_{M,d_{n}}(x, 0)=\delta(x)
\end{eqnarray}
\end{subequations}
we can find the Fourier-Laplace transforms $\widehat{P}_{M,a_{n}}(k, s)$ and $\widehat{P}_{M,d_{n}}(k,s)$ according to
\begin{subequations}
\begin{eqnarray}
& \widehat{P}_{M,a_{n}}(k,s)=\frac{1}{s}\left[\prod_{i=1}^M\left(\frac{r_{i}}{s+r_{i}+D k^{2}}\right)^{m_{i}}\right], \\
& \widehat{P}_{M,d_{n}}(k, s)=\frac{1}{s+D k^{2}} \left[1-\prod_{i=1}^M\left(\frac{r_{i}}{s+r_{i}+D k^{2}}\right)^{m_{i}}\right]  .
\end{eqnarray}
\end{subequations}
As a sanity check, we can easily verify that $P_{M,d_n}(x,t)$ integrates to $1$ at $t=0$ and to $0$ as $t\to\infty$ and that $\int P_{M,a_n}\,\mathrm{d}x\to 1$ as $t\to\infty$, confirming probability conservation.

For $P_{M,a_{n}}$, we can use the same partial fraction decomposition as equation (\ref{dec}):
\begin{equation}
\label{eq:PMa_decomp}
P_{M,a_{n}}(x, t)=\sum_{i=1}^M \sum_{j=1}^{m_{i}} \frac{A_{i, j}}{r_{i}^{j}} P_{r_i,a_{j}}\left(x, t \right).
\end{equation}
Similarly for $P_{M,d_{n}}$, we find
\begin{equation}
\label{eq:PMd_decomp}
P_{M,d_{n}}(x, t)=G_{0}(x, t)+\sum_{i=1}^M \sum_{j=1}^{m_{i}} \frac{A_{i, j}}{r_{i}^{j}}\left(P_{r_i,d_{j}}\left(x, t  \right)-G_{0}(x, t)\right) .
\end{equation}
equations~(\ref{eq:PMa_decomp}) and~(\ref{eq:PMd_decomp}) are the degenerate generalisations of the non-degenerate decompositions~(\ref{PraND}) and~(\ref{PrdND}): in the limit $m_i\to 1$ for all $i$, the double sums reduce to single sums and the $A_{i,j}$ coefficients reduce to the Lagrange weights of~(\ref{PraND}).

\subsection{Spatial moments via the characteristic functions}

Since the steady-state of particle $n$ for heterogeneous resetting rates can be written as a weighted sum of single particle distributions,
the spatial moments can be expressed in terms of single particle moments. For example, in the case of non-degenerate rates, we can write
\begin{equation}
\left\langle x_{n}^{2 q}\right\rangle=\sum_{j=1}^{n}\left\langle x_{1}^{2 q}\right\rangle_{r=r_{j}} \prod_{i=1, i \neq j}^{n} \frac{r_{i}}{r_{i}-r_{j}}.
\end{equation}
While we can rewrite the sum along similar lines to equation (\ref{ssnondeg2}), it is more straightforward to work with the more general degenerate expression (\ref{sso}):
\begin{equation}
\widetilde{P}_{M,n}^{*}(k)=\prod_{i=1}^{M}\left(\frac{r_{i}}{r_{i}+D k^{2}}\right)^{m_{i}}=\prod_{i=1}^{M} \widetilde{P}_{r_i,m_{i}}^{*}\left(k\right),  
\label{eq:PMn}
\end{equation}
which is a product of $M$ characteristic functions, each corresponding to an $m_i$-particle equal-rate chain with rate $r_i$. Equivalently, in real space, $P_{M,n}^*(x)$ is the convolution of $M$ such distributions. The Fourier transform $\widetilde{P}_{M,n}^*(k)$ is the \emph{characteristic function} of the steady-state distribution of particle $n$. Spatial moments are obtained from it by differentiation at $k=0$:
\begin{equation}
\left\langle x_n^{2q}\right\rangle
= (-1)^q\left.\frac{\mathrm{d}^{2q}\widetilde{P}_{M,n}^*(k)}
  {\mathrm{d}k^{2q}}\right|_{k=0}.
\end{equation}

The $2q$-th moment then follows from the Leibniz rule for the $2q$-th derivative of a product of $M$ functions, evaluated at $k=0$:
\begin{equation}
\left.\frac{\mathrm{d}^{2 q} \widetilde{P}_{M,n}^{*}(k)}{\mathrm{d} k^{2 q}}\right|_{k=0}=\left.\sum_{k_{1}+k_{2}+\cdots+k_{M}=2 q}\binom{2 q}{k_{1}, k_{2}, \ldots, k_{M}} \prod_{i=1}^{M} \frac{\mathrm{d}^{k_{i}} \widetilde{P}_{r_i,m_{i}}^{*}\left(k \right)}{\mathrm{d} k^{k_{i}}}\right|_{k=0} 
\end{equation}
where the sum runs over all integers such that $\sum_{i=1}^{M} k_{i}=2 q $ and the multinomial coefficient is
\begin{equation}
\binom{2 q}{k_{1}, k_{2}, \ldots, k_{M}}=\frac{(2 q)!}{k_{1}!k_{2}!\ldots k_{M}!} .
\end{equation}

Since only even-order derivatives of $\widetilde{P}_{r_i,m_i}^*(k)$ are non-zero at $k=0$ (by the symmetry $P_{r_i,m_i}^*(x)=P_{r_i,m_i}^*(-x)$), the sum reduces to
\begin{equation}
\left.\frac{\mathrm{d}^{2q}\widetilde{P}_{M,n}^{*}(k)}
{\mathrm{d}k^{2q}}\right|_{k=0}
= \sum_{\ell_1+\cdots+\ell_M=q}
  \binom{2q}{2\ell_1,2\ell_2,\ldots,2\ell_M}
  \prod_{i=1}^{M}
  \left.\frac{\mathrm{d}^{2\ell_i}\widetilde{P}_{r_i,m_i}^{*}(k)}
  {\mathrm{d}k^{2\ell_i}}\right|_{k=0}.
\end{equation}
Note that this is consistent with the fact that only even-order spatial moments are non-zero by symmetry. Recasting this equation in terms of spatial moments gives
\begin{align}
(-1)^{q}\left\langle x_{n}^{2 q}\right\rangle &=\sum_{\ell_{1}+\ell_{2}+\cdots+\ell_{M}=q}\binom{2 q}{2 \ell_{1}, 2 \ell_{2}, \ldots, 2 \ell_{M}} \prod_{i=1}^{M}(-1)^{\ell_{i}}\left\langle x_{m_{i}}^{2 \ell_{i}}\right\rangle_{r=r_{i}}  \nonumber \\
\fl &=(-1)^{q} \sum_{\ell_{1}+\ell_{2}+\cdots+\ell_{M}=q}\binom{2 q}{2 \ell_{1}, 2 \ell_{2}, \ldots, 2 \ell_{M}} \prod_{i=1}^{M}\left\langle x_{m_{i}}^{2 \ell_{i}}\right\rangle_{r=r_{i}} . 
\end{align}
Hence, we arrive at the final result
\begin{equation}
\label{xn1}
\left\langle x_{n}^{2 q}\right\rangle=\sum_{\ell_{1}+\ell_{2}+\cdots+\ell_{M}=q}\binom{2 q}{2 \ell_{1}, 2 \ell_{2}, \ldots, 2 \ell_{M}} \prod_{i=1}^{M}\left\langle x_{m_{i}}^{2 \ell_{i}}\right\rangle_{r=r_{i}}, 
\end{equation}
which produces an alternative form of the moments which allows us to easily take the limit $r_{i} \rightarrow r_{j}$. We thus conclude that the $2q$-th moment of particle $n$ is a multinomial convolution of the $2\ell_i$-th moments of the individual rate-blocks: each block $i$ contributes $m_i$ particles resetting at rate $r_i$, and the total moment is obtained by summing over all ways of distributing the total order $2q$ among the $M$ blocks, weighted by the multinomial coefficient. This formula allows the limit $r_i\to r_j$ to be taken without difficulty, since each factor $\langle x_{m_i}^{2\ell_i}\rangle_{r=r_i}$ is a smooth function of $r_i$.

For $q=1$,  equation (\ref{xn1}) gives the variance for non-degenerate rates
\begin{equation}
\left\langle x_{n}^{2}\right\rangle=\sum_{i=1}^{M} \frac{2 D m_i}{r_{i}} 
\label{xn2}
\end{equation}
which recovers the constant case by setting $M=1$ and $m_1=n$, see equation (\ref{xn}). Further, for $q=2$, we find
\begin{align}
\label{xn4}
\left\langle x_{n}^{4}\right\rangle&=\sum_{i=1}^{M}\left(\frac{12 D^{2} m_i (m_i + 1)}{r_{i}^{2}}+3 \sum_{j \neq i}^{M} \frac{2 D m_i}{r_{i}} \cdot \frac{2 D m_j}{r_{j}}\right)\\
&=12 D^{2}\left[\sum_{i=1}^{M} \frac{m_i}{r_{i}^{2}} + \left( \sum_{i=1}^{M} \frac{m_i}{r_{i}}\right)^{2}\right]
\end{align}
allowing access to the excess kurtosis.

\begin{figure}[t!]
\begin{center}
\includegraphics[width=\linewidth]{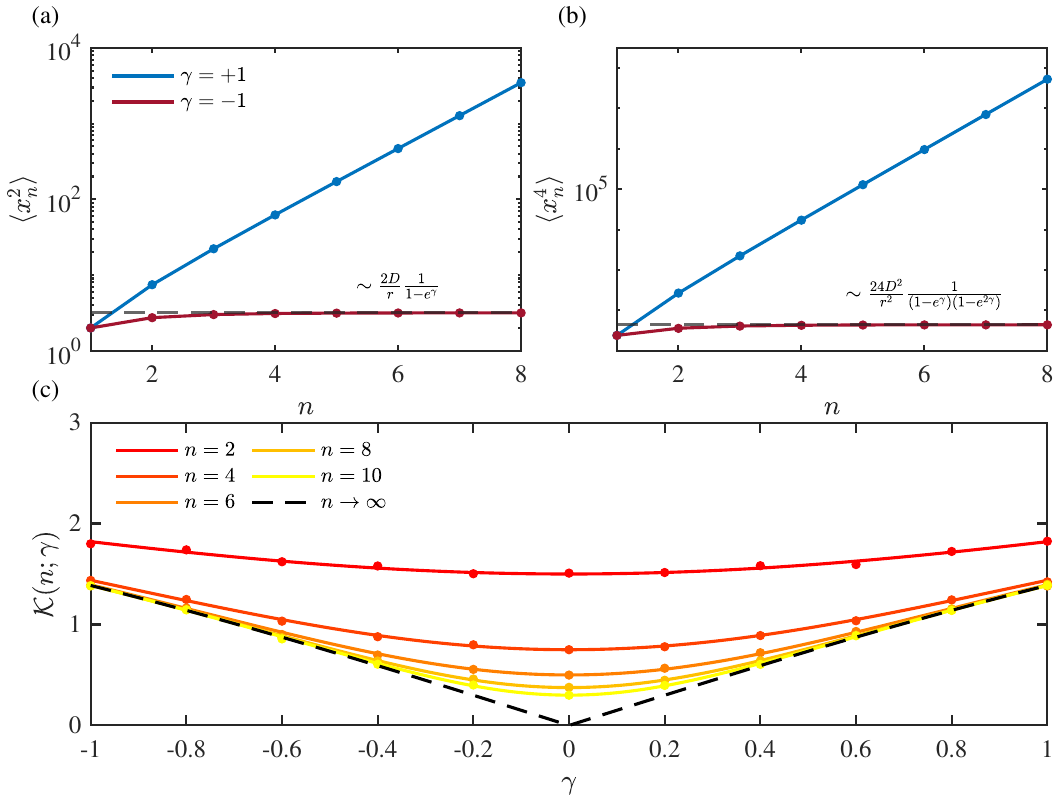}
\caption{\textit{Steady-state properties of nested resetters with geometric rates.} (a) Steady-state spatial variance $\langle x^2_n \rangle$ of nested resetters with geometrically decreasing (red) and increasing (blue) resetting rates, equation (\ref{geometricvar}) $(r_i = re^{-\gamma(i-1)},$ with $r=1, D=1)$; (b) Steady-state fourth spatial moment $\langle x^4_n \rangle$ of nested resetters with geometrically decreasing (red) and increasing (blue) resetting rates, equation (\ref{geometricfourth}) $(r_i = re^{-\gamma(i-1)},$ with $r=1,D=1)$; (c) Excess kurtosis $\mathcal{K}(n;\gamma)$ of nested resetters with geometric resetting rates as a function of $\gamma$, equation (\ref{Kurt2}), plotted for various $n$ with $r=1,D=1$. Solid lines indicate analytical results and symbols indicate numerical simulations.}
\label{fig:geometricmoment}
\end{center}
\end{figure}

\subsection{Examples of heterogeneous resetting rates}

\subsubsection{Geometric resetting rates.}
As our first example, we consider geometric resetting rates as an example of non-degenerate rates, with $r_i=re^{-\gamma(i-1)}$ for $i=1,\ldots,n$. Starting from equation~(\ref{xn2}), we find for the variance
\begin{equation}\label{geometricvar}
\left\langle x_{n}^{2}\right\rangle =\frac{2 D}{r} \sum_{i=1}^{n} e^{\gamma(i-1)}   
 =\frac{2 D}{r} \frac{e^{\gamma n}-1}{e^{\gamma}-1} 
\end{equation}
For $\gamma>0$, we have an exponentially increasing variance. For $\gamma<0$, the variance saturates as $n \rightarrow \infty$
\begin{equation}
\lim _{\mathbf{n} \rightarrow \infty} \frac{2 D}{r} \frac{e^{\gamma n}-1}{e^{\gamma}-1}=\frac{2 D}{r} \frac{1}{1-e^{\gamma}} . 
\end{equation}
Finally, for $\gamma=0$, we recover the equal resetting rate variance (\ref{xn}) via L'Hôpital's rule:
\begin{equation}
\lim _{\gamma \rightarrow 0} \frac{2 D}{r}   \frac{e^{\gamma n}-1}{e^{\gamma}-1}=\frac{2 n D}{r} .  \end{equation}
For the fourth moment, we find
\begin{equation}\label{geometricfourth}
\left\langle x_{n}^{4}\right\rangle=\frac{12 D^{2}}{r^{2}}\left[\left(\frac{e^{\gamma n}-1}{e^{\gamma}-1}\right)^{2}+\frac{e^{2 \gamma n}-1}{e^{2 \gamma}-1}\right]  .
\end{equation}
We once again see exponential growth for $\gamma>0$ and saturation for $\gamma<0$ with
\begin{equation}
\lim _{n \rightarrow \infty} \left\langle x_{n}^{4}\right\rangle =\frac{24 D^{2}}{r^{2}\left(1-e^{\gamma}\right)\left(1-e^{2 \gamma}\right)} .  
\end{equation}
The excess kurtosis reads
\begin{equation}
\label{Kurt2}
\mathcal{K}(n)=3 \cdot \frac{e^{2 \gamma n}-1}{e^{2 \gamma}-1} \cdot\left(\frac{e^{\gamma}-1}{e^{\gamma n}-1}\right)^{2} .  \end{equation}
For $\gamma=0$, this recovers $\mathcal{K}(n)=3/n$ (see equation~(\ref{Kurt})). Curiously, for $\gamma \neq 0$, the excess kurtosis (\ref{Kurt2}) converges to a non-zero value as $n \rightarrow \infty$
\begin{equation}
\lim_{n \to \infty}\mathcal{K}(n)=\frac{3\left|e^{\gamma}-1\right|}{e^{\gamma}+1}=3\tanh \left(\frac{|\gamma|}{2}\right) .
\end{equation}
The excess kurtosis interpolates between $\mathcal{K}=0$ (Gaussian, $\gamma=0$) and $\mathcal{K}=3$ (Laplace, $|\gamma|\to\infty$). The interpolation between Gaussian and Laplace limits has a clear physical interpretation. For $\gamma=0$, all particles reset at the same rate and the steady-state distribution of particle $n$ is the $n$-fold convolution of a Laplace distribution; by the CLT this approaches a Gaussian as $n\to\infty$, giving $\mathcal{K}\to 0$, \cite{Alston25b}. For $|\gamma|\to\infty$, one particle's dynamics dominate entirely: for $\gamma<0$ (increasing rates) particle $n$ resets fastest and dominates, while for $\gamma>0$ (decreasing rates) particle $1$ dominates; in either case the distribution approaches that of a single Laplace-distributed resetter, giving $\mathcal{K}\to 3$. 

The symmetry in $|\gamma|$ arises because the characteristic function~(\ref{sso}) is a product of single-particle characteristic functions, so the order of the resetting rates along the chain is irrelevant for the steady-state statistics of particle $n$. This means that a geometrically decreasing system can be mapped to a geometrically increasing system and vice versa (with appropriately adjusted prefactors, the value of which the kurtosis is agnostic to)
\begin{align}
    r_i = r e^{-\gamma(i-1)} \mapsto r_{i}' = \bigl(re^{-\gamma(n-1)}\bigr)e^{\gamma(i-1)},
\end{align}
where now particle $i$ resets with rate $r_i' = r_{n-i+1}$. 

\subsubsection{Periodic resetting rates.}
Next we consider once again periodic resetting rates as an example of degenerate rates. The variance (\ref{xn2}) is  
\begin{equation}\label{periodicvar}
\left\langle x_{n}^{2}\right\rangle=2 D \biggl(m \sum_{i=1}^{M} \frac{1}{r_{i}}+\sum_{j=1}^{(n-m M)} \frac{1}{r_{j}}\biggr) .
\end{equation}
For large $n, m \simeq n / N$ and the second term is $\mathcal{O}(1)$ so that
\begin{equation}\label{reff}
\left\langle x_{n}^{2}\right\rangle \simeq \frac{2 D n}{r_{\mathrm{eff}}} ,\quad r_{\mathrm{eff}}=\frac{M}{\sum_{i=1}^{M} \frac{1}{r_{i}}},
\end{equation}
where $r_{\mathrm{eff}}$ is the \emph{harmonic mean} of the resetting rates. The variance thus grows linearly in $n$ with an effective rate set by the harmonic mean, which is dominated by the \emph{slowest} resetters in the chain. 

\begin{figure}[t!]
\begin{center}
\includegraphics[width=\linewidth]{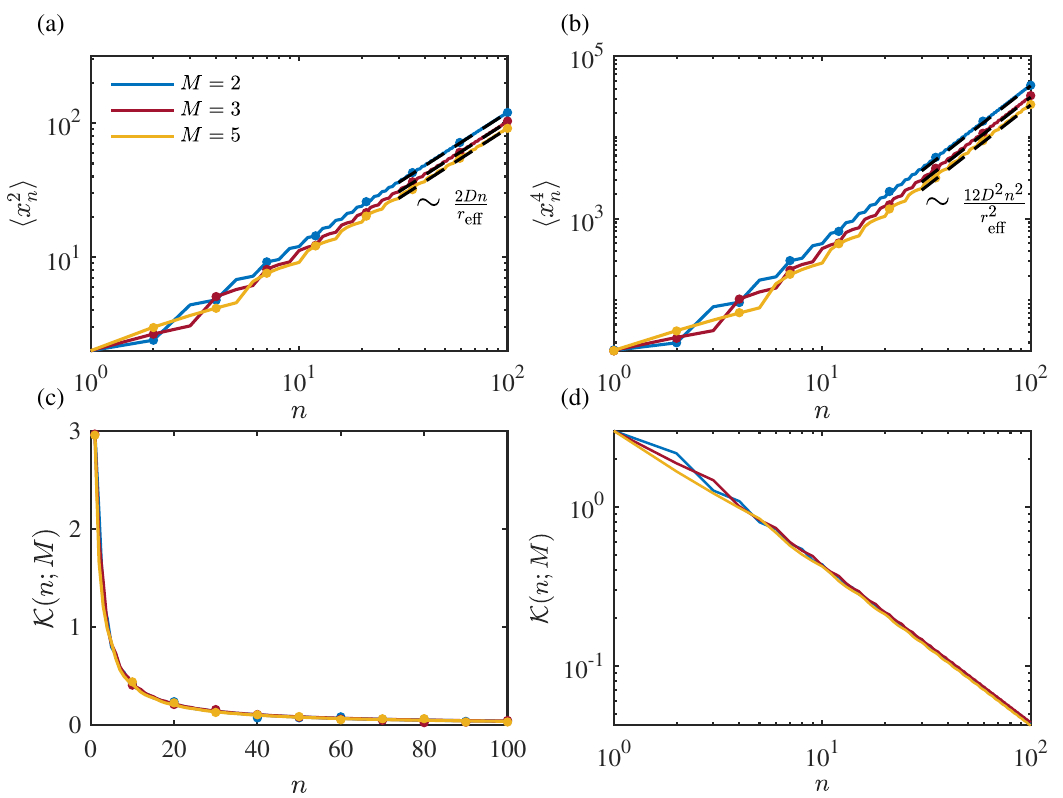}
\caption{\textit{Steady-state properties of nested resetters with periodic rates.} (a) Steady-state spatial variance  $\langle x^2_n \rangle$ of nested resetters with periodic resetting rates of varying period size $M$, equation (\ref{periodicvar}), where increasing $M$ increases the relative size of plateaus seen in the curve; (b) Steady-state fourth spatial moment $\langle x^4_n \rangle$ of nested resetters with periodic resetting rates of varying period size $M$, equation (\ref{periodicfourth}), where increasing $M$ once again increases the relative size of plateaus. (c) Excess kurtosis $\mathcal{K}(n;M)$ of nested resetters with periodic resetting rates as a function of $n$, equation (\ref{Kurt3}), plotted with periodic resetting rates of varying period size $M$ with inset showing the same data on a log-log scale. For all observables the resetting rates are linearly spaced between $1$ and $5$, and $D=1$. Solid lines indicate analytical results, dashed lines indicate large $n$ behaviour (\ref{reff}), and symbols indicate numerical simulations.}\label{fig:periodicmoments}
\end{center}
\end{figure}

Further, the fourth moment (\ref{xn4}) is 
\begin{equation}\label{periodicfourth}
\left\langle x_{n}^{4}\right\rangle=12 D^{2}\left[m \sum_{i=1}^{M} \frac{1}{r_{i}^{2}}+\sum_{j=1}^{(n-m M)} \frac{1}{r_{j}^{2}}+\left(m \sum_{i=1}^{M} \frac{1}{r_{i}}+\sum_{j=1}^{(n-m M)} \frac{1}{r_{j}}\right)^{2}\right],
\end{equation}
where once again large $n$ behaviour is found to be $\langle x_n^4\rangle\simeq 12D^2n^2/r_{\mathrm{eff}}^2$.
The excess kurtosis is finally given by
\begin{equation}\label{Kurt3}
\mathcal{K}(n)=\frac{3\left[m \sum_{i=1}^{M} \frac{1}{r_{i}^{2}}+\sum_{j=1}^{(n-m M)} \frac{1}{r_{j}^{2}}\right]}{\left(m \sum_{i=1}^{M} \frac{1}{r_{i}}+\sum_{j=1}^{(n-m M)} \frac{1}{r_{j}}\right)^{2}}  
\end{equation}
which approaches zero as $n \rightarrow \infty$, consistent with the Gaussian limit identified above. We numerically confirm all results in Figure \ref{fig:periodicmoments}.

\subsection{Accumulation times for heterogeneous resetting rates}

We now extend the analysis of accumulation times in section 3 to nested particles with heterogeneous resetting rates. 

\subsubsection{Accumulation times for two particles with different rates.}
First consider a pair of particles with resetting rates $\r=(r_1,r_2)$, $r_1\neq r_2$. equations (\ref{PraND}) and (\ref{PrdND}) become
\begin{subequations}
\begin{align}
P_{\r,a_{2}(x,t)} &=P_{r_1,a_{1}}\left(x, t \right) \frac{r_{2}}{r_{2}-r_{1}}-P_{r_2,a_{1}}\left(x, t \right) \frac{r_{1}}{r_{2}-r_{1}}, \\
P_{\r,d_{2}(x,t)} &=P_{r_1,d_{1}}\left(x, t \right) \frac{r_{2}}{r_{2}-r_{1}}-P_{r_2,d_{1}}\left(x, t  \right) \frac{r_{1}}{r_{2}-r_{1}}.
\end{align}
\end{subequations}
The fractional deviation from the NESS for particle $2$ now reads
\begin{equation}
Z_{\r,a_{2}}(x,t)=1-\frac{P_{\r,a_{2}}(x, t)}{P_{\r,2}^{*}(x)}=\frac{\left(r_{2} e^{-r_{1} t} P_{r_1,1}^{*}\left(x \right)-r_{1} e^{-r_{2} t} P_{r_2,1}^{*}\left(x \right)\right) * G_{0}(x, t)}{r_{2} P_{r_1,1}^{*}\left(x \right)-r_{1} P_{r_2,1}^{*}\left(x  \right)} .
\end{equation}
The accumulation time of particle 2 is then
\begin{align}
T_{\r,a_{2}}(x) & =\frac{\frac{r_{2}}{r_{1}} P_{r_1,1}^{*}\left(x \right) * P_{r_1,1}^{*}\left(x\right)-\frac{r_{1}}{r_{2}} P_{r_2,1}^{*}\left(x  \right) * P_{r_2,1}^{*}\left(x \right)}{r_{2} P_{r_1,1}^{*}\left(x  \right)-r_{1} P_{r_2,1}^{*}\left(x  \right)}  \nonumber\\
& =\frac{|x|}{\sqrt{4 D}} \cdot \frac{r_{2} \exp \left(-\alpha_{1}|x|\right)\left(1+\frac{1}{\alpha_{1}|x|}\right)-r_{1} \exp \left(-\alpha_{2}|x|\left(1+\frac{1}{\alpha_{2}|x|}\right)\right.}{r_{2} \sqrt{r_{1}} \exp \left(-\alpha_{1}|x|\right)-r_{1} \sqrt{r_{2}} \exp \left(-\alpha_{2}|x|\right)} .
\end{align}
If $\alpha_1^{-1}\gg\alpha_2^{-1}$, i.e.\ $r_1\ll r_2$ (particle~$1$ resets much more slowly than particle~$2$), then at large $|x|$ the terms involving $e^{-\alpha_1|x|}$ dominate over those involving $e^{-\alpha_2|x|}$, since $\alpha_1\ll\alpha_2$. In this limit, $T_{\mathbf{r},a_2}(x)$ takes the form of a single-particle accumulation time governed by the slower rate $r_1$:
\begin{equation}
\label{eq:T2_het_approx}
T_{\r,a_{2}}(x) \sim \frac{|x|+\alpha_{1}^{-1}}{\sqrt{4 D r_{1}}}.
\end{equation}
which anticipates the general result of Section~\ref{sec:accum_het_N}: the accumulation time of the full chain
is controlled by the particle with the minimum resetting rate.

\subsubsection{\texorpdfstring{Accumulation times for $N$-particles nested with non-degenerate rates.}{Accumulation times for N-particles nested with non-degenerate rates.}}
\label{sec:accum_het_N}
We now extend our $N$-particle result to particles with different rates; we first assume that all rates are non-degenerate. The fractional deviation takes the form
\begin{equation}
\label{ZnND}
Z_{\r,a_{n}}(x, t)=\sum_{i=1}^{n}\left[\prod_{j=1, j \neq i}^{n} \frac{r_{j}}{r_{j}-r_{i}}\right] \frac{P_{r_{i},1}^{*}\left(x  \right) * e^{-r_{i} t} G_{0}(x, t)}{P_{\r,n}^{*}(x)} .
\end{equation}
We find the accumulation time of particle $n$ to be
\begin{eqnarray}
T_{\r,a_{n}}(x) & =\sum_{i=1}^{n} \frac{P_{r_{i},1}^{*}\left(x  \right) * P_{r_{i},1}^{*}\left(x \right)}{r_{i} P_{\r,n}^{*}(x)} \prod_{j=1, j \neq i}^{N} \frac{r_{i}}{r_{j}-r_{i}}  \nonumber\\
& =\frac{\sum_{i=1}^{n} \frac{1}{r_{i}} P_{r_{i},1}^{*}\left(x \right) * P_{r_{i},1}^{*}\left(x  \right) \prod_{j=1, j \neq i}^{n} \frac{r_{j}}{r_{j}-r_{i}}}{\sum_{i=1}^{n} P_{r_{i},1}^{*}\left(x  \right) \prod_{j=1, j \neq i}^{n} \frac{r_{j}}{r_{j}-r_{i}}}.
\end{eqnarray}
which, upon substituting the explicit form of $P_{r_\ell,1}^*(x)$, gives
\begin{equation}\label{eq:tRan}
T_{\r,a_{n}}(x)=\frac{|x|}{\sqrt{4 D}} \cdot \frac{\sum_{i=1}^{n} \exp \left(-\alpha_{i}|x|\right)\left(1+\frac{1}{\alpha_{i}|x|}\right) \prod_{j=1, j \neq i}^{n} \frac{r_{j}}{r_{j}-r_{i}}}{\sum_{i=1}^{n} \sqrt{r_{i}} \exp \left(-\alpha_{i}|x|\right) \prod_{j=1, j \neq i}^{n} \frac{r_{j}}{r_{j}-r_{i}}}, 
\end{equation}
where $\alpha_{i}^{-1}=\sqrt{D / r_{i}}$ is the length scale associated with particle $i$. 

As our calculation has assumed particle $i$ has a unique resetting rate $r_{i}$ (and therefore unique length scale $\alpha_{i}^{-1}$ ), there necessarily is a particle with a minimum resetting rate $r_{\text {min }}$ (equivalently $\alpha_{\min }^{-1}$ ) such that $r_{\min }\leq  r_{i}$ for all $i$ (equivalently $\alpha_{\min }^{-1} \geq \alpha_{i}^{-1}$ ). The large $|x|$ behaviour of~(\ref{eq:tRan}) is governed by the condition that the exponentials $e^{-\alpha_i|x|}$ are well-separated in scale. Specifically,
the approximation below requires both:
\begin{enumerate}
\item $|x|\gg\alpha_{\min}^{-1}$ (so that the accumulation time is in the large $|x|$ regime), and
\item $\alpha_{\min}\ll\alpha_i$ for all $i\neq i^*$ (a separation of scales in the resetting rates, i.e.\ $r_{\min}\ll r_i$ for all $i\neq i^*$).
\end{enumerate}
When condition~(2) holds, the terms $e^{-\alpha_i|x|}$ for $i\neq i^*$ are exponentially smaller than $e^{-\alpha_{\min}|x|}$ at large $|x|$, and the sums in~(\ref{eq:tRan}) are dominated by the $i=i^*$ term. When two rates $r_i$ and $r_j$ are close but not equal, condition~(2) is violated for the pair $(i,j)$. In this regime, both exponentials contribute comparably and the accumulation time interpolates between the single-rate result~(\ref{eq:T2_het_approx}) and the minimum-rate result~(\ref{Tmin}). The crossover occurs on the length scale $|x|\sim(\alpha_i-\alpha_j)^{-1}=(\sqrt{r_j/D}-\sqrt{r_i/D})^{-1}$, beyond which the minimum-rate term dominates. For rates that are nearly degenerate ($r_i\approx r_j$), this crossover length can be very large, and the degenerate rates results should be used instead.

\begin{figure}[t!]
\begin{center}
\includegraphics[width=\linewidth]{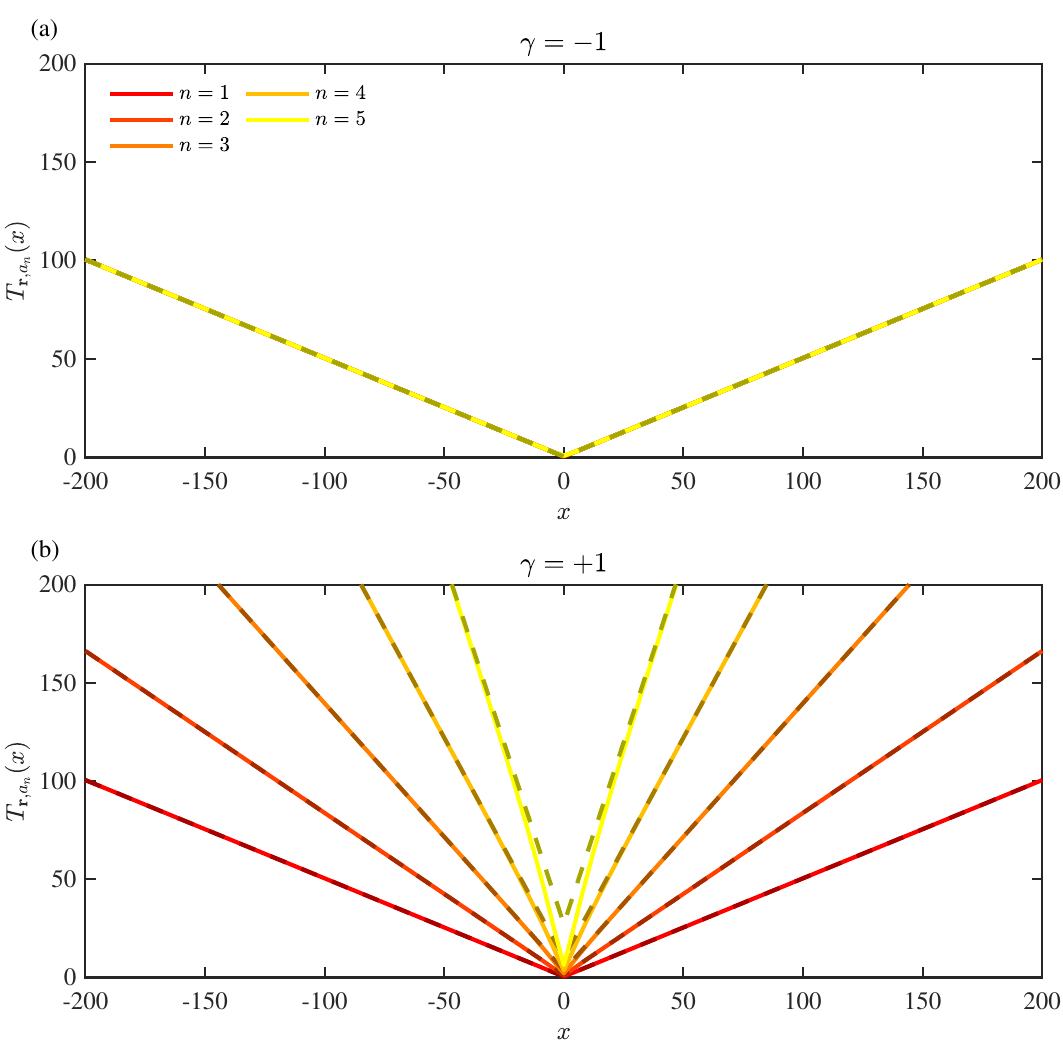}
\caption{\textit{Accumulation time $T_{\mathbf{r},a_n}(x)$ of nested resetters with geometric rates.} Accumulation times of nested resetters with (a) geometrically increasing resetting rates $(r_i = e^{i-1})$ and (b) geometrically decreasing resetting rates $(r_i = e^{1-i})$ both with diffusivity $D=1$. Solid lines indicate analytical results, equation~(\ref{eq:tRan}); dashed lines indicate large $|x|$ behaviour, equation~(\ref{Tmin}).}\label{fig:geometrictran}
\end{center}
\end{figure}

Under conditions~(1) and~(2), the sums in~(\ref{eq:tRan}) are dominated by the $i=i^*$ term, giving
\begin{equation}
T_{\r,a_{n}}(x)  \sim \frac{|x|}{\sqrt{4 D}} \cdot \frac{\exp \left(-\alpha_{\min }|x|\right)\left(1+\frac{1}{\alpha_{\min }|x|}\right)}{\sqrt{r_{\min }} \exp \left(-\alpha_{\min }|x|\right)} =\frac{|x|+\alpha_{\min }^{-1}}{\sqrt{4 D r_{\min }}} .
\label{Tmin}
\end{equation}
This implies that, in assuming that there is a particle such that $\alpha_{\text {min }} \ll \alpha_{i}$, the approach to steady state is controlled almost entirely by the minimum resetting rate on the chain. We confirm that indeed equation (\ref{Tmin}) captures equation (\ref{eq:tRan}) at large $|x|$ in Figure \ref{fig:geometrictran}. In fact, this is exactly equivalent to stating that the position of particle $n$ at steady state can be approximated by the position of a single-particle stochastic resetter with resetting rate $r_{\text {min}}$. Equivalently, the steady-state distribution of particle $n$ is approximated by a single-particle Laplace distribution with rate $r_{\min}$:
\begin{equation}
P_{\r,n}^{*}(x) \sim P_{r_{\min},1}^{*}\left(x  \right) . 
\end{equation}
Applying this idea to equation (\ref{tan}) one can view all the particles in a standard nested resetting chain as having the same minimum resetting rate, which then results in proportionality to $n$, the number of particles with minimum resetting rate $r$, as all particles on average diffuse the same distance between resets.

\subsubsection{\texorpdfstring{Accumulation times for $N$-particles nested with degenerate rates.}{Accumulation times for N-particles nested with degenerate rates.}}
If we instead allow for degeneracy, bridging the gap between our two results, equations (\ref{Zn}) and (\ref{ZnND}), we find that the fractional deviation takes the form
\begin{equation}
\label{eq:ZMan}
Z_{M,a_n}(x,t) = \frac{\displaystyle\sum_{i=1}^M\sum_{j=1}^{m_i} \frac{A_{i,j}}{r_i^j} \sum_{k=0}^{j-1}\frac{(r_i t)^k}{k!}\, e^{-r_i t}\,P_{r_i,j-k}^{*}(x)*G_0(x,t)} {\displaystyle\sum_{i=1}^M\sum_{j=1}^{m_i} \frac{A_{i,j}}{r_i^j}\,P_{r_i,j}^{*}(x)},
\end{equation}
and the accumulation time is
\begin{equation}
\label{eq:TMan}
T_{M,a_n}(x) = \frac{\displaystyle\sum_{i=1}^M\sum_{j=1}^{m_i} \frac{jA_{i,j}}{r_i^{j+1}}\, P_{r_i,j}^{*}(x)*P_{r_i,1}^{*}(x)} {\displaystyle\sum_{i=1}^M\sum_{j=1}^{m_i} \frac{A_{i,j}}{r_i^j}\,P_{r_i,j}^{*}(x)}.
\end{equation}

Writing the PDFs explicitly, we can argue that the dominant terms in the numerator and denominator of~(\ref{eq:TMan}) at large $|x|$ are those with the minimum resetting rate and the highest multiplicity. This requires the same separation-of-scales condition as in Section~\ref{sec:accum_het_N}: $\alpha_{\min}\ll \alpha_i$ for all $i\neq i_{\rm min}$, where $i_{\rm min}$ is the index of the particle with the minimum resetting rate. We then only keep the terms $i=i_{\rm min}$ and $j=m_{i_{\rm min}}$ in the double sums. Hence, setting $r_{\min }=r_{i_{\rm min}}$ and $m_{\min}=m_{i_{\rm min}}$, we have
\begin{equation}\label{eq:tRan2}
T_{M,a_{n}}(x) \sim \frac{m_{\min} P_{r_{\min},m_{\min}}^{*}\left(x \right) * P_{r_{\min},1}^{*}\left(x \right)}{r_{\min } P_{r_{\min},m_{\min }}^{*}\left(x \right)}, 
\end{equation}
where the $A_{i, j}$ prefactor cancels between numerator and denominator. Expanding for $|x| \gg \left(m_{\text {min }}-1\right) \alpha_{\text {min }}^{-1}$ using the large $|x|$
Bessel function expansion~(\ref{aBessel}), we arrive at
\begin{equation}\label{eq:tMin2}
T_{M,a_{n}}(x) \sim \frac{|x|+m_{\min } \alpha_{\min }^{-1}}{\sqrt{4 D r_{\min }}}, 
\end{equation}
which confirms that indeed the approach to steady state is governed by both the minimum resetting rate in the nested chain, $r_{\text {min }}$, \emph{and} the number of particles with said resetting rate, $m_{\min }$. Comparing with~(\ref{Tmin}), the effect of degeneracy is to replace the single delay $\alpha_{\min}^{-1}$ by $m_{\min}\alpha_{\min}^{-1}$, consistent with the equal-rate result~(\ref{tan}) in which the delay is $n\alpha^{-1}$ for a chain of $n$ particles all resetting at rate $r$. 

\begin{figure}[t!]
\begin{center}
\includegraphics[width=\linewidth]{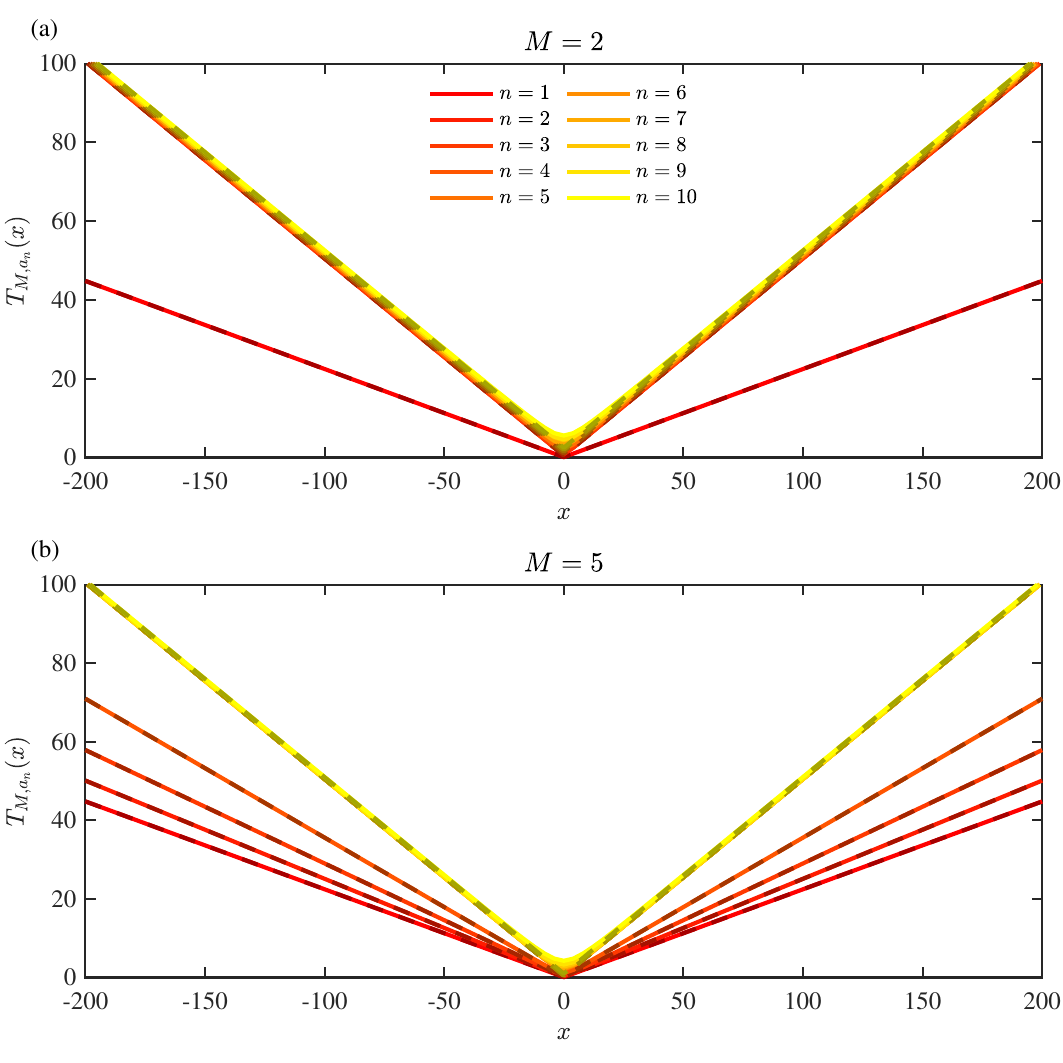}
\caption{\textit{Accumulation time of nested resetters with periodic rates.} Accumulation times $T_{M,a_n}(x)$ of nested resetters with periodic resetting rates and diffusivity $D=1$: (a) $\mathbf{r} = (5,1)$;(b) $\mathbf{r} = (5,4,3,2,1)$. Solid lines indicate analytics, equation (\ref{eq:tRan2}), and dashed lines indicate large $|x|$ behaviour, equation (\ref{eq:tMin2}).}\label{fig:periodictran}
\end{center}
\end{figure}

Our result is thus equivalent to approximating the position of particle $n$ at steady state as the position of particle $m_{\text {min }}$ in a chain of particles with equal resetting rate $r_{\text {min }}$ at large $|x|$, i.e. $x \gg\left(m_{\text {min }}-1\right) \alpha_{\text {min }}^{-1}$, leading to the following steady-state distribution
\begin{equation}
P_{n}^{*}(x) \sim P_{r_{\min},m_{\min }}^{*}\left(x \right) . 
\end{equation}
We confirm that indeed equation (\ref{eq:tMin2}) captures equation (\ref{eq:tRan2}) at large $x$ in Figure \ref{fig:periodictran}. 

\section{Discussion and conclusion}
\label{sec:conclusion}

In this work, we have provided a comprehensive analytical characterisation of the approach to nonequilibrium steady state in nested stochastic resetting processes, extending the steady-state analysis of Ref.~\cite{Alston25b} to the full time-dependent regime for a general model including heterogeneous resetting rates. We began by deriving the full time-dependent probability distribution for a nested resetting chain with equal rates (equation~(\ref{eq:full_PDF})), obtained by decomposing the distribution into an accumulating component $P_{r,a_n}(x,t)$, which converges to the NESS $P_{r,n}^*(x)$, and a decaying component $P_{r,d_n}(x,t)$, which vanishes as $t\to\infty$. As already seen in Ref.~\cite{Alston25b}, we confirm that the steady-state distribution is a modified Bessel function of the second kind (equation~(\ref{ssn})), whose spatial moments (equation~(\ref{xn})) reveal an approach to Gaussianity along the chain, quantified by the excess kurtosis $\mathcal{K}(n)=3/n$.

We then characterised the approach to the NESS via the accumulation time $T_{r,a_n}(x)$, the effective timescale for the local establishment of steady state at position $x$, following the method developed in Ref.~\cite{Bressloff21}. For a single particle, the accumulation time takes the form $T_{r,a_1}(x)=(|x|+\alpha^{-1})/\sqrt{4Dr}$ (equation~(\ref{T1a})), which can be interpreted as the time for a characteristic wavefront, propagating with effective speed $\sqrt{4Dr}$, to reach position $x$, with an inherent delay $\alpha^{-1}/\sqrt{4Dr}$ reflecting the initial spreading of the distribution over the reset length scale $\alpha^{-1}$. For $N$ particles with equal rates, our key result is that this delay increases linearly with the particle index $n$: $T_{r,a_n}(x)\sim(|x|+n\alpha^{-1})/\sqrt{4Dr}$. This linear growth reflects the sequential propagation of information down the resetting chain, with each particle contributing an additive delay of $\alpha^{-1}/\sqrt{4Dr}$ before it can establish the NESS of the next particle in the hierarchy.

Furthermore, we extended the analysis to heterogeneous resetting rates. In the non-degenerate case, the time-dependent distribution is a linear combination of single-particle distributions with Lagrange-basis weights (equation~(\ref{Pnhet})), arising from the partial-fraction decomposition of the renewal equation in the Laplace variable. In the degenerate case, the Fokker-Planck equation yields distributions that are weighted sums of $m_i$-particle equal-rate distributions, where $m_i$ is the multiplicity of each distinct rate (equation~(\ref{dec})). The spatial moments of the heterogeneous chain are given by a multinomial convolution formula (equation~(\ref{xn1})). We focused on two examples. First, for a periodic chain with period $M$, the variance grows as $\langle x_n^2\rangle\simeq 2Dn/r_{\mathrm{eff}}$ for large $n$, where $r_{\mathrm{eff}}=M/\sum_{i=1}^M r_i^{-1}$ is the harmonic mean of the resetting rates; the variance is thus bottlenecked by the slowest resetters in the chain. Second, for geometric rates $r_i=re^{-\gamma(i-1)}$, the excess kurtosis saturates as $n\to\infty$ to $3\tanh(|\gamma|/2)$, interpolating between the Gaussian limit ($\gamma=0$) and the Laplace limit ($|\gamma|\to\infty$). For the accumulation times with heterogeneous rates, relaxation to the NESS is governed by the minimum resetting rate $r_{\min}$ in the chain and, in the degenerate case, by its multiplicity $m_{\min}$: $T_{M,a_n}(x)\sim(|x|+m_{\min}\alpha_{\min}^{-1})/ \sqrt{4Dr_{\min}}$ (equation~(\ref{eq:tMin2})). This bottleneck result unifies the equal-rate case (where $m_{\min}=n$) and the non-degenerate case (where $m_{\min}=1$).

The nested resetting chain is an example of a many-body nonequilibrium model that admits exact, closed-form solutions for the full time-dependent distribution, all spatial moments, and the accumulation times characterising relaxation to the NESS. This tractability stems from two structural features: the renewal property of Poissonian resetting, which allows the time-dependent distribution to be expressed as a convolution of the propagator with the reset distribution via the last-renewal equation, and the nested convolution structure of the steady-state distribution, which reduces the many-body problem to a sequence of single-particle problems connected by the partial-fraction decomposition of the characteristic function.

The present analysis involves a number of simplifying assumptions, each of which could be relaxed in future work. First, the analysis is restricted to one spatial dimension, where the Gaussian propagator and Laplace steady-state distribution take particularly simple forms; generalisation to $d>1$ is interesting but non-trivial. Second, we have assumed Poissonian resetting throughout; for non-Poissonian protocols such as power-law or deterministic resetting, the partial-fraction structure of the time-dependent distribution is lost, though the accumulation time framework should remain applicable since it relies only on the decomposition $P=P_a+P_d$, and the single-particle case has been studied in~\cite{Bressloff22}. Third, we have set $P_0(x)=\delta(x)$; for a non-trivial reset distribution the Bessel function structure of $P_{r,n}^*$ is modified, but the renewal equation framework and the definition of the accumulation time remain valid. 

Several further extensions of this work present themselves naturally. The steady-state pairwise correlations $\langle X_n X_{n+j}\rangle$ were computed in Refs.~\cite{Alston25b}; the time-dependent correlations $\langle X_n(t)X_{n+j}(t+\tau)\rangle$ during relaxation to the NESS have not been computed here and would be a natural extension of the accumulation time framework. The mean first-passage time to a target at position $L$ for the $n$-th particle is another natural quantity: based on the bottleneck argument of Section~\ref{sec:accum_het_N}, we expect it to be governed by $r_{\min}$ in the heterogeneous case and to display a linear dependence on $n$ in the equal-rate case, analogous to the linear delay in the accumulation time. Moving from a linear chain to a general directed graph (for instance a tree structure in which a root particle drives multiple downstream branches) would be a significant generalisation relevant to biological signalling cascades; the Fokker-Planck approach extends immediately to any graph topology since the steady-state recursion generalises to a system of linear equations on the graph. Finally, the entropy production rate of the nested chain has not been computed here; for a single resetting particle this quantity is known to be related to the Kullback-Leibler divergence between the instantaneous distribution and the NESS~\cite{Bressloff22}, and extending this to the nested setting would complement the kinetic results derived in the present work.
\section*{Acknowledgements}
C.B. was supported by a Roth Scholarship.
\appendix

\section{\texorpdfstring{Proof of equation\,\ref{renn2} by induction}{Proof of equation 2.10 by induction}}
\label{app:induction}

In the main text, we have showed that the PDF for particle $n$ could be written in a closed form as 
\begin{equation}
\label{renn2_app}
P_{r,n}(x,t) = e^{-rt}\sum_{j=0}^{n-1}\frac{(rt)^{j}}{j!}\,G_{0}(x,t) + r\int_{0}^{t}\mathrm{d}\tau\,\frac{(r\tau)^{n-1}}{(n-1)!}\, e^{-r\tau}\,G_{0}(x,\tau).
\end{equation}
This result can be established by induction on $n$ as we will show here. First, recall the Chapman--Kolmogorov identity for the Gaussian propagator
\begin{equation}
\label{eq:CK_app}
G_0(x,t) = \int_{-\infty}^{+\infty}\mathrm{d}\xi\; G_{0}(x-\xi,\tau)\,G_0(\xi,t-\tau),
\end{equation}
which expresses the semigroup property of the free-diffusion propagator.

The base case $n=1$ is equation~(\ref{ren1})
\begin{equation}
\label{ren1_app}
P_{r,1}(x,t) = e^{-rt}\,G_{0}(x,t) + r\int_0^t \mathrm{d}\tau\, e^{-r\tau}\,G_{0}(x,\tau).
\end{equation}
which holds by inspection.

Assume that~(\ref{renn2_app}) holds for $n-1$, i.e.\
\begin{equation}
\label{eq:inductive_hypothesis}
P_{r,n-1}(x,t) = e^{-rt}\sum_{j=0}^{n-2}\frac{(rt)^{j}}{j!}\,G_{0}(x,t) + r\int_{0}^{t}\mathrm{d}\tau\,\frac{(r\tau)^{n-2}}{(n-2)!}\, e^{-r\tau}\,G_{0}(x,\tau).
\end{equation}
Substituting~(\ref{eq:inductive_hypothesis}) into the convolution term of the renewal equation
\begin{equation}
P_{r,n}(x,t) = e^{-rt}\,G_{0}(x,t) + r\int_0^t \mathrm{d}\tau\, e^{-r\tau} \int^{+\infty}_{-\infty}\mathrm{d} \xi \;G_{0}(x-\xi,\tau)\,P_{r,n-1}(\xi,t-\tau),
\label{renn_app}
\end{equation}
gives
\begin{align}
r\int_0^t \mathrm{d}\tau\, &e^{-r\tau} \int_{-\infty}^{+\infty}\mathrm{d}\xi\;G_{0}(x-\xi,\tau)\,P_{r,n-1}(\xi,t-\tau) = r\int_0^t \mathrm{d}\tau\, e^{-r\tau} \int_{-\infty}^{+\infty}\mathrm{d}\xi\; G_{0}(x-\xi,\tau)  \nonumber \\
&\times \Bigg[ e^{-r(t-\tau)}\sum_{j=0}^{n-2}\frac{(r(t-\tau))^{j}}{j!}\,G_{0}(\xi,t-\tau) + r\int_{0}^{t-\tau}\mathrm{d}s\,\frac{(rs)^{n-2}}{(n-2)!}\,e^{-rs}\,G_{0}(\xi,s)
\Bigg]. \label{eq:inductive_step}
\end{align}
We treat the two terms in~(\ref{eq:inductive_step}) separately.

\textit{Term 1.} Using the Chapman--Kolmogorov identity~(\ref{eq:CK_app}), the spatial integral collapses to $G_0(x,t)$, and
the first term becomes
\begin{align}
r\int_0^t \mathrm{d}\tau\, e^{-rt}
\sum_{j=0}^{n-2}\frac{r^j(t-\tau)^{j}}{j!}\,G_{0}(x,t)
&= r\,e^{-rt}\,G_0(x,t)
\sum_{j=0}^{n-2}\frac{r^j}{j!}
\int_0^t (t-\tau)^j\,\mathrm{d}\tau \notag \\
&= r\,e^{-rt}\,G_0(x,t)
\sum_{j=0}^{n-2}\frac{r^j}{j!}\cdot\frac{t^{j+1}}{j+1} \notag \\
&= e^{-rt}\,G_0(x,t)
\sum_{j=1}^{n-1}\frac{(rt)^{j}}{j!},
\label{eq:term1}
\end{align}
where in the last step we substituted $j\to j-1$.

\textit{Term 2.} Applying the Chapman--Kolmogorov identity~(\ref{eq:CK_app}) to the spatial integral and exchanging the order of integration in $\tau$ and $s$, the second term becomes
\begin{align}
r^2\int_0^t \mathrm{d}\tau\, e^{-r\tau}
\int_0^{t-\tau}\mathrm{d}s\,\frac{(rs)^{n-2}}{(n-2)!}\,e^{-rs}\,G_0(x,\tau+s).
\end{align}
Setting $u = \tau + s$ and exchanging the order of integration:
\begin{align}
&= r^2\int_0^t \mathrm{d}u\, G_0(x,u)
\int_0^{u}\mathrm{d}\tau\,
e^{-r\tau}\,\frac{r^{n-2}(u-\tau)^{n-2}}{(n-2)!}\,e^{-r(u-\tau)} \notag \\
&= r\int_0^t \mathrm{d}u\, e^{-ru}\,G_0(x,u)
\cdot r^{n-1}\int_0^{u}\mathrm{d}\tau\,
\frac{(u-\tau)^{n-2}}{(n-2)!} \notag \\
&= r\int_0^t \mathrm{d}u\,\frac{(ru)^{n-1}}{(n-1)!}\,
e^{-ru}\,G_0(x,u).
\label{eq:term2}
\end{align}
Adding~(\ref{eq:term1}) and~(\ref{eq:term2}) to the no-reset term $e^{-rt}G_0(x,t)$ (the $j=0$ contribution) from~(\ref{renn_app}), we recover exactly~(\ref{renn2}), completing the induction.

\section*{References}
 \bibliographystyle{iopart-num}
 \bibliography{references}

\end{document}